\documentclass{article}
\usepackage[utf8]{inputenc}

\usepackage{arxiv}
\usepackage[T1]{fontenc}

\usepackage{booktabs}
\usepackage[flushleft]{threeparttable}
\usepackage[numbers]{natbib}
\RequirePackage[colorlinks,linkcolor=blue, citecolor=blue,urlcolor=blue]{hyperref}
\usepackage{amsmath,amssymb}          
\usepackage{graphicx}				
\usepackage{psfrag,epsf}
\usepackage{epstopdf}								
\usepackage{booktabs}
\usepackage{multirow}
\usepackage{siunitx}
\usepackage{caption}
\usepackage{tikz}
\usepackage{float}
\usepackage{ragged2e}
\hypersetup{colorlinks,linkcolor={blue},citecolor={blue},urlcolor={blue}}
\usepackage{color}
\usetikzlibrary{shapes.geometric, arrows}
\usepackage{lineno}

\newcommand{\con}{{\,|\,}}

\newcommand{\LL}{{L(\theta_1,\theta_0 ; n)}}
\newcommand{\Ln}{{L_n}}
\DeclareMathOperator*{\argminA}{arg\,min}

\tikzstyle{io} = [trapezium, trapezium left angle=70, trapezium right angle=110, minimum width=3cm, minimum height=.8cm, text centered, text width=4cm, draw=black, fill=blue!30]
\tikzstyle{process} = [rectangle, minimum width=3cm, minimum height=.1cm, text centered, text width=4cm, draw=black, fill=orange!30]
\tikzstyle{decision} = [diamond, minimum width=.1cm, minimum height=.1cm, text centered, text width=2.5cm, draw=black, fill=green!30]
\tikzstyle{arrow} = [thick,->,>=stealth]

\title{A Modified Sequential Probability Ratio Test}

\author{
	Sandipan Pramanik\\
	Department of Statistics\\
	Texas A\&M University\\
	\texttt{sandy.pramanik@gmail.com}\\
	\And
	Valen E. Johnson\thanks{Corresponding author.}\\
	Department of Statistics\\
	Texas A\&M University\\
	\texttt{vjohnson@stat.tamu.edu}\\
	\And
	Anirban Bhattacharya\\
	Department of Statistics\\
	Texas A\&M University\\
	\texttt{anirbanb@stat.tamu.edu}\\
}

\begin{document}

	\large
	
	\maketitle
	
	\begin{abstract}
		We describe a modified sequential probability ratio test that can be used to reduce the average sample size required to perform statistical hypothesis tests at specified levels of significance and power. Examples are provided for $z$ tests, $t$ tests, and tests of binomial success probabilities. A description of a software package to implement the test designs is provided. We compare the sample sizes required in fixed design tests conducted at 5\% significance levels to the average sample sizes required in sequential tests conducted at 0.5\% significance levels, and we find that the two sample sizes are approximately equal.
	\end{abstract}
	

\section{Introduction}
\begin{sloppypar}
	Experimental science relies on controlled experiments that test whether effects predicted by a scientific theory can be produced and measured in laboratory settings. Observational science is based on measuring outcomes as they occur naturally, without experimental intervention. In practice, measured outcomes from both observational studies and experiments are subject to random variation and measurement error. For this reason, hypothesis testing procedures must be employed to determine whether data support or do not support a hypothesized effect. In the classical hypothesis testing paradigm, two types of errors are considered when making this assessment.  Type I error 
	occurs when the null hypothesis of ``no effect'' is rejected when
	the hypothesized effect does not exist.   To limit claims of false discovery, hypothesis testing procedures are commonly designed so that the probability of a Type I error (i.e., $\alpha$) is limited to be less than a prespecified value, often 0.05.  Type II error 
	occurs when we fail to reject the null hypothesis when the hypothesized effect does exist (the probability of a Type II error is denoted by $\beta$).

	Recent concerns over the replicability of scientific studies have led to calls to move away from $p$ values and significance testing \citep{amrhein2019,blakeley2019,pike2019,savalei2015}. However, $p$ values and significance testing continue to play critical roles in many areas, including genomics, high-energy physics, and clinical trials. An examination of recent articles in prominent psychology journals also suggests that $p$ values and significance testing continue to play an important role in psychological research \citep{OSC2015,johnson17}. Elsewhere, we have proposed to address the limitations of $p$ values by reducing the significance thresholds required for declaring a positive finding from $\alpha=0.05$ to $\alpha=0.005$ \citep{benjamin18, johnson13_1}.  While this change would improve the replicability of scientific claims of discoveries, it would also increase the costs of conducting studies because larger sample sizes would be required if similar controls on Type II error 
	probability were maintained.

	This article describes a modification of the sequential probability ratio test (SPRT) of \cite{wald45} that reduces the sample sizes required to achieve specified Type I and Type II error 
	probabilities.  The modified design can be applied to many studies conducted in the social and natural sciences in which the goal is to  establish the existence of a hypothesized effect. Implicit in this goal is the detection of effects that are not arbitrarily close to zero (or the null value of the parameter).  In this regard, the proposed design differs from recent developments of sequential procedures designed to estimate various effect sizes, such as standardized mean differences, correlation and regression coefficients, and coefficients of variation \citep{chat16,kelley18,kelley19}.

	\textcolor{black}{We propose the Modified Sequential Probability Ratio Test (MSPRT) for testing a point null hypothesis against a one or two-sided alternative hypothesis. In designing these tests, we objectively set alternative hypotheses. The alternative hypotheses we propose are based on uniformly most powerful Bayesian tests (UMPBT's) or approximate UMPBT's \citep{johnson13_1,johnson13_2}. Details regarding UMPBT's appear in Section~\ref{sec:Modified SPRT Method}. We note that exact UMPBT's are known only for one-parameter exponential family models and tests for the non-centrality parameters of chi-squared statistics \citep{nikooienejad2020}. Approximate UMPBT's are known for {\em t} tests.  Thus, a limitation of the MSRPT is that is applicable primarily to {\em z} and {\em t} tests, tests of binomial proportions and Poisson means, and chi-squared tests.}

	\textcolor{black}{For this class of tests,  empirical evidence suggests that MSPRT's require sample sizes that can be less than 50\% of the sample size that is required in corresponding fixed designs when the null hypothesis of no effect is true, and sample sizes that can be 20\% smaller when alternative hypotheses are true.  In general, the sample size savings accrued by the use of the MSPRT depends on the test statistic chosen and the targeted Type I and II error probabilities for the test.  Empirical studies illustrating such savings are described in Section~\ref{sec:Simulation}. Theoretical support for these findings is provided in \cite{sieg1985}, where approximate formulae for the average sample number (ASN) and operating characteristics for truncated SPRTs are derived. These results approximate discrete time stochastic processes (representing the observed sequential tests) by Brownian motion or Wiener processes, which are continuous time stochastic processes. For sufficiently large sample sizes, these processes provide approximate operating characteristics and ASN's for truncated SPRT's. In the case of one- and two-sample $z$ tests, the underlying assumptions required in deriving those formulae apply to the MSPRT, and the approximate values from these results agree with our empirical findings. Specific details regarding this connection appear in Section~\ref{sec:Performance in one-sample tests}. As noted by \cite{sieg1985}, this theoretical result ``leads to appreciable qualitative insight; and quantitatively it does provide a first, crude approximation which can often be used as a basis for subsequent refinement.''   
	}
	
	\textcolor{black}{The remainder of this article is organized as follows. Section~\ref{sec:Sequential Testing Procedures} reviews sequential hypothesis testing procedures. In Section~\ref{sec:Modified SPRT Method}, we define MSPRT's, and in Section~\ref{sec:Implementation} we describe R code that can be used to implement them. In Section~\ref{sec:Simulation} we present numerical findings from simulation studies, and compare the performance of the MSPRT to group sequential designs and sequential Bayes factors \citep{schon17}. Section~\ref{sec:realdata} complements Section~\ref{sec:Simulation} by applying the MSPRT to the gambler fallacy study data \citep{oppenheimer2009} collected in the Many Labs~1 project \citep{klein2014}. Finally, we summarize our findings in Section~\ref{sec:Discussion}.
	}

\end{sloppypar}

\section{Sequential Testing Procedures}\label{sec:Sequential Testing Procedures}

\begin{sloppypar}
	In contrast to fixed sample size 
	designs, sequential testing procedures provide a rule for stopping a study after observing individual  
	participants or groups of participants.   A sequential testing procedure specifies a rule that decides, after a group of participants has been measured, whether to (i) continue to collect data, (ii) stop data collection and reject the null hypothesis, or (iii) stop 
	data collection and reject the alternative hypothesis.

	Sequential testing procedures have not previously found widespread application in behavioral and social science research.  However, the statistical theory for these tests has been developed extensively since their introduction by Wald in the 1940s. For a comprehensive review of statistical theory underlying these procedures, see \cite{sieg1985}. Most applications have occurred  
	in item response theory (IRT) and computer adaptive test designs, where sequential tests are often used to terminate IRT-based adaptive classification tests \citep{king83, eggen99}.   Other recent applications include an item selection algorithm in a binary IRT model \citep{nydick14} and an extension to Bayesian hypothesis testing, called ``Sequential Bayes Factors," that provides an optional stopping rule for multiple testing \citep{schon17}. From a theoretical point of view, a bound for the expected stopping time (i.e., the test length) was obtained in adaptive mastery tests for dependent data \citep{chang04}.

	The SPRT is one of the most widely known sequential testing procedures \citep{wald45,lai2001,lai2004,lai2008,bartroff2008,bar2018}. This \textcolor{black}{test} 
	is based on comparing the likelihood ratio between a simple (i.e., point or precise) null hypothesis and a simple alternative hypothesis, and stopping 
	data collection as soon as the likelihood ratio strongly supports one of the two.

	To illustrate this procedure in more detail, suppose that independent data values are collected sequentially. Denote these values by $x_1,x_2,\dots$ .  Suppose further that the null hypothesis implies that the probability density function describing a single data value $x_i$ is $f(x_i\con \theta_0)$, and that the alternative hypothesis implies that the probability density function is $f(x_i \con \theta_1)$.
	Then the likelihood ratio in favor of the alternative hypothesis based on the first $n$ observations is defined as
	\begin{equation}\label{LR}
	\LL = \prod_{i=1}^n \frac{ f(x_i ; \theta_1)}{f(x_i ; \theta_0)}.
	\end{equation}
	To simplify notation, we denote $\LL$ by $\Ln$.

	Heuristically, the SPRT keeps track of the likelihood ratio $L_n$ as data accumulate, and stops the experiment as soon as the probability assigned to the data under one hypothesis significantly exceeds the probability assigned to the data by the other hypothesis.

	More formally, the SPRT proceeds by comparing $\Ln$, $n=1,2,\dots$, to constants $A$ and $B$, $A>B>0$, as data from individual 
	study participants are collected.  The procedure stops when 
	$\Ln \geq A$ or $\Ln \leq B$, 
	or equivalently when $\Ln$ exits the interval $(B,A)$ for the first time.  The quantities $A$ and $B$ are defined as 
	\begin{equation}\label{wald_boundary}
	A = \frac{1-\beta}{\alpha} \qquad \mbox{and} \qquad B = \frac{\beta}{1-\alpha}.
	\end{equation}
	If $\Ln \geq A$, the null hypothesis is rejected; if $\Ln \leq B$, 
	the alternative hypothesis is rejected.
	An important property of the SPRT is that it requires, on average, fewer participants to achieve its specified Type I and Type II error probabilities than any other test 
	whose error probabilities are smaller than or the same as these \citep{wald1948}.

	A key limitation of the SPRT is that it requires the specification of both a null hypothesis and an alternative hypothesis.  Specifying an alternative hypothesis is not required in classical hypothesis tests when only Type I error probability constraints have been imposed. The proposed MSPRT addresses this limitation by implicitly deriving the alternative hypothesis from the design parameters according to pre-specified criteria. From a user's point of view, this eliminates the need to explicitly specify an  alternative hypothesis, even though the procedure does, of course, directly depend on the alternative hypothesis that is used. For this reason, users should carefully consider the magnitude of the effect size implicit in the MSPRT to determine whether it represents a plausible alternative hypothesis.  In this regard, the use of the MSPRT mimics classical experimental design procedures in which Type I and II error 
	probabilities, sample size, and targeted effect size  are balanced against each other to determine a suitable test design.

	Another limitation of the SPRT is that the sample size required to complete a test cannot be determined prior to the start of data collection.   In nearly all experimental settings, resources available for testing participants are limited and in observational studies the amount of the data that can be collected from a population is finite. 
	This feature of the SPRT thus complicates the practical design of 
	tests and is resolved by the MSPRT. An earlier modification of the SPRT, known as the truncated SPRT, was proposed by \cite{anderson60} to address this difficulty. However, this modification generally provides less statistical power than our proposed MSPRT. For instance,  Tables 3.1 and 3.11 in \cite{sieg1985} indicate that for the alternative effect size that provides 80\% power in a fixed design test, the truncated SPRT's power is only 74\%. By comparison, the MSPRT provides between 78-79\% power at the same alternative. Further examples of this difference are provided in Section~\ref{sec:Performance in one-sample tests}, where we describe similar differences in power for other effect sizes.

	Modifications of the SPRT proposed to handle composite hypotheses are primarily of two types. One is known as the weighted SPRT and was proposed by \cite{wald45}. This test replaces the likelihood ratio with the ratio of integrated likelihoods, weighted with respect to given weight functions for the respective hypotheses. The weight functions are determined by losses associated with incorrectly accepting various alternative hypotheses. The other type of modification is known as the generalized SPRT, which is based on the ratio of maximized likelihoods under the respective hypotheses, and is similar to the generalized likelihood ratio (GLR) test \citep{lai1991}.

	Other extensions of the SPRT, the MaxSPRT and the sequential GLR test, were proposed for drug and vaccine safety surveillance \citep{kulldorf,shih2010}. The goal of the MaxSPRT is to reject the null hypothesis (that a treatment is safe) if there is substantial evidence that a treatment is not safe. Like the SPRT, the MaxSPRT does not impose a bound on the maximum sample size $N$. In addition, the design does not allow early rejection of the alternative hypothesis. The sequential GLR was proposed to address these issues. Importantly, both tests are based on the GLR, and the alternative hypothesis used by them is the maximum likelihood estimate (MLE) of the parameter being tested. When the null hypothesis is true, the MLE converges to the null value, and as a consequence the tests never terminate a trial in favor of the null hypothesis. Furthermore, for sufficiently large $N$ the tests can, in principle, reject null hypotheses for arbitrarily small effect sizes.

	From a practical perspective, there are many hypothesis testing contexts where it is not feasible to implement a SPRT. For instance, a 
	SPRT cannot be applied 
	when data are not collected sequentially.  Similarly, it cannot be applied when it is not possible to perform the evaluation as soon as participants are treated.  Such is the case in clinical trials of new disease therapies, which are often conducted at multiple treatment centers.   Collation of data across centers can be time consuming, and it can be difficult to convene review boards. In addition, patient outcomes are often not known for months or even years after a treatment has been administered.  To address these challenges, group sequential designs have been developed to allow for the evaluation of patient outcomes only after groups of patients have been observed or at scheduled interim analysis times \citep{landemet94, hay71, jennturn99, obf79, peto85,  pocock77}. \cite{sieg86, sieg1985} provides detailed discussion of the termination of repeated significance tests for group sequential studies with a maximum sample size.
\end{sloppypar}

\section{The Modified SPRT}\label{sec:Modified SPRT Method}

\begin{sloppypar}
	To address the limitations of existing sequential tests, we propose a modified SPRT (MSPRT) in which 
	
	\begin{itemize}
		\item  the maximum sample size ($N$) required in a hypothesis test is 
		fixed prior to the start of an experiment, and 
		\item  the effect size defining the alternative hypothesis and used to sequentially compute the likelihood ratio $L_n$ 
		is derived from the size of the test $\alpha$ (Type I error probability), the maximum available sample size $N$, and the targeted Type II error probability, $\beta$. 
	\end{itemize}
	Thus, $N$, $\alpha$, and $\beta$ are MSPRT design parameters that are fixed at the outset of the study.  The effect size defining the alternative hypothesis is determined from these values. Given these values, the MSPRT is defined in a manner similar to Wald's initial proposal.
	
	To objectively set the alternative hypothesis in the MSPRT, we find the uniformly most powerful Bayesian test (UMPBT) or the approximate UMPBT that matches the rejection region of a classical test of size $\alpha$ with a sample size of $N$ \citep{johnson13_2}. Under fixed designs, UMPBT's are tests that maximize the probability that the Bayes factor in favor of the alternative hypothesis exceeds a specified threshold over the class of all alternative hypotheses. \cite{johnson13_2} showed that such tests can be obtained by assuming a point alternative and then maximizing the probability mentioned above with respect to such alternatives. The optimum value of the point alternative is defined as the UMPBT alternative. We defer a more detailed description of UMPBTs to Sections S1--S3 of the supplemental materials. The key feature of an UMPBT relevant to our purpose is that it provides an automated procedure for defining an alternative hypothesis against which the null hypothesis is tested.
	For sampling densities that belong to the class of one-parameter exponential family models (including $z$ tests, tests for proportions, and tests of means of Poisson counts), UMPBTs exist.  For other sampling densities, and in particular for $t$ tests, approximate UMPBTs exist.  In many cases, the values of the parameter that define the alternative hypotheses in these tests are approximately equal to the maximum likelihood estimate of the parameter obtained from data that lie on the boundary of the rejection region of the test.  
	
	To illustrate a simple UMPBT, consider a size $\alpha$ $z$ test of $H_0 : \theta = \theta_0$ versus $H_1 : \theta > \theta_0$ based on $N$ samples from a normal population with an unknown mean $\theta$ and known standard deviation $\sigma$. For this problem, the UMPBT alternative hypothesis is $\theta= \theta_0 + z_\alpha \sigma/\sqrt{N}$, where $z_\alpha$ is the $100(1-\alpha)$th quantile of a standard normal distribution. 

	Table \hyperlink{table}{1} provides the UMPBT alternatives that can be used in some common, one-sided null hypothesis significance tests.  In this table, definitions of alternative hypotheses are determined by 
	the maximum sample size $N$ for the $z$ test and test of a binomial proportion. For the $t$ test, it also depends on $n$, the currently observed sample size. These alternatives are used to compute the likelihood ratio at each step. Thus for a $t$ test, the alternative hypothesis used to compute the likelihood ratio changes after each data point is collected and a new estimate of the observational variance is obtained. For the $z$ and $t$ tests, the UMPBT alternatives are point alternatives.  The alternative for the test of a binomial proportion is a mixture distribution of two proportions; a mixture density is used to achieve more accurate Type I error probability control due to 
	the discrete nature of the binomial distribution.
	
	To understand the nature of this mixture distribution, it is necessary to introduce additional notation.  Denote the cumulative distribution function (cdf), inverse cdf, and the probability mass function of a binomial distribution with denominator $N$ and success probability $\theta$ by $F(\cdot; N, \theta)$, $\bar{F}(\cdot; N, \theta)$, and $f(\cdot; N, \theta)$, respectively. Given $N$ and $\alpha$ for a right one-sided test of the probability $\theta$, define the cut-off point $c_0$ in a fixed design test by
	\[ c_0 = \inf \Big \{ c = 0, 1, \cdots, N \Big| \bar{F}(c; N, \theta_0 ) \leq \alpha \Big \}.\]
	For $\theta \in [0,1]$ and $\delta>0$, let
	\[ h_N (\theta, \delta) =  \frac{ \log \delta - N \bigg[ \log(1-\theta)- \log(1- \theta_0) \bigg]}{\displaystyle{ \log\bigg(\frac{\theta}{1-\theta}\bigg) - \log\bigg(\frac{\theta_0}{1- \theta_0}}\bigg)},\]
and define $\theta (\delta) = \argminA_{\theta>\theta_0} \, h_N (\theta, \delta)$. With these ingredients, we define the UMPBT alternative as the mixture distribution
	\[\theta \sim \psi_R \, I_{\theta = \theta_{R,L}} + (1-\psi_R ) \, I_{\theta = \theta_{R,U}},\]
	where $I_{a=b}$ is 1 if $a=b$ and 0 otherwise.
	Also, $\theta_{R,L} = \theta \big( \delta_{R,L} \big)$  and $\theta_{R,U} = \theta \big( \delta_{R,U} \big)$, where $\delta_{R,L}$ and $\delta_{R,U}$ satisfy
	\[h_N \Big(\theta \big( \delta_{R,L} \big), \delta_{R,L} \Big) = c_0 -1, \,\, \mbox{and} \,\, h_N \Big( \theta \big( \delta_{R,U} \big), \delta_{R,U} \Big) = c_0 ,\]
	and $\psi_R = [\alpha - \bar{F}(c_0 ; N, \theta_0 )]/f(c_0 ; N, \theta_0 )$. A similar derivation can be applied to left one-sided tests. 
	Further details are provided in \cite{johnson13_1,johnson13_2} and Section~S3 in the supplemental document.
	
	In practice, of course, researchers should examine the design parameters of a MSPRT before the sequential design is initiated.  That is, the alternative hypothesis generated by the MSPRT in order to obtain the targeted Type I and Type II error probabilities should be inspected, as should the actual error probabilities achieved by the test design.  If the implied effect size is either unreasonably large or substantively unimportant, then investigators should reconsider the maximum sample size and error probability controls that were specified.  

	{\color{black}
	
	In the case of one-sided hypothesis testing, given the alternative hypothesis obtained from the UMPBT or approximate UMPBT, Wald's SPRT is conducted either until the likelihood ratio ($z$ and $t$ tests) or the weighted likelihood ratio (proportion test) exits the interval $(B,A)$ or until $N$ samples (e.g., study participants) have been tested.
	The values of $A$ and $B$ for the MSPRT are the same as those used in Wald's test and, as noted previously, are given by
	\[  A = \frac{1-\beta}{\alpha} \qquad \mbox{and} \qquad B = \frac{\beta}{1-\alpha} .\]
	
	If no decision has been reached after exhausting $N$ samples, a threshold $\gamma$ is determined numerically so that the Type I error probability of the test equals $\alpha$ for continuous data and is less than or equal to $\alpha$ for discrete data.  If $L_N \geq \gamma$, the null hypothesis $H_0$ is rejected and the experiment is terminated. Otherwise, if $L_N < \gamma$, the alternative hypothesis $H_1$ is rejected and the experiment is terminated.
	
	The extension of the MSPRT for two-sided tests is accomplished by simultaneously running two one-sided tests of size $\alpha/2$. Before reaching the maximum sample size $N$, the test terminates by $(a)$ rejecting $H_0$ when either of the tests reject $H_0$, or $(b)$ by not rejecting $H_0$ if both the tests reject $H_1$. If the test continues to the maximum sample size $N$, then a common termination threshold, $\gamma$, is determined so as to maintain the desired Type I error probability of the test. The design parameter $\gamma$ is chosen to be as small as possible while still maintaining the specified size of the test, $\alpha$. If $L_N \geq \gamma$ for either of the tests, the null hypothesis is rejected. Otherwise, the test rejects the alternative hypothesis.
	
	}
	
	In practice, it may be useful to examine the value of $L_n$ at the termination of an MSPRT.  This value represents the likelihood ratio between hypotheses based on all accumulated data and may be of particular interest when a test terminates after the maximum sample size has been reached. \textcolor{black}{Of course, an advantage of formal hypothesis testing procedures is that they encourage investigators to design experiments that have a reasonably high probability of providing ``signficant'' evidence in favor of a scientifically important effect.  At the end of a parametric hypothesis test, it is usually possible to compute the likelihood ratio in favor of the MLE over the null parameter value.  In the particular case of the MSPRT, the investigator is able to go a step further and report the Bayes factor in favor of an alternative hypothesis which was considered scientifically acceptable before the experiment was undertaken. The MSPRT thus encourages the design of tests that will lead to Bayes factors (or likelihood ratios) that differ substantially from 1.0, and they do so with smaller sample sizes than are required in fixed design tests. The values of the likelihood ratio $L_n$ are provided by the {\bf \fontfamily{qcr} \selectfont MSPRT} software described \textcolor{black}{in Section~\ref{sec:Implementation}}.}

	\begin{table}
	\centering
	\begin{threeparttable}
		\caption{\em UMPBT alternatives for one-sided tests}\label{umpbt_table}
		\begin{tabular}{c c c c}
				\midrule[1.3pt]
				Test & $H_0$	&	$H_1$ & UMPBT alternative \\ [0.5ex]
				\midrule
				
				\multirow{2}*{$z$ test}	& \multirow{2}*{$\theta=\theta_0$} & $\theta > \theta_0$	& $\theta = \theta_0 + z_{\alpha} \frac{\sigma}{\sqrt{N}}$	\\
				\cmidrule (l){3-4}
				
				& & $\theta < \theta_0$	& $\theta = \theta_0 - z_{\alpha} \frac{\sigma}{\sqrt{N}}$	\\ 			\midrule
				
				\multirow{2}*{$t$ test}	& \multirow{2}*{$\theta=\theta_0$} & $\theta > \theta_0$	& $\theta = \theta_0 + t_{\alpha; N-1} \frac{s_{n}}{\sqrt{N}}$	\\
				\cmidrule (l){3-4}
				
				& & $\theta < \theta_0$	& $\theta = \theta_0 - t_{\alpha; N-1} \frac{s_{n}}{\sqrt{N}}$	\\ \midrule
				
				\multirow{2}*{Test for proportion}	& \multirow{2}*{$\theta=\theta_0$} & $\theta > \theta_0$	& $\theta \sim \psi_R \, I_{\theta = \theta_{R,L}} + (1-\psi_R ) \, I_{\theta = \theta_{R,U}}$\\
				\cmidrule (l){3-4}
				
				& & $\theta < \theta_0$	& $\theta \sim (1-\psi_L ) \, I_{\theta = \theta_{L,L}} + \psi_L \, I_{\theta = \theta_{L,U}}$\\
				\midrule[1.3pt]
		\end{tabular}
		\begin{tablenotes}
		\small
		\item {\em Note.} For one-sample $z$ and $t$ tests, UMPBT alternative hypotheses have closed-form expressions. For one-sample tests of proportions, (non-randomized) MSPRT's can be used to more accurately achieve Type I error probability control, but a mixture distribution is required as the alternative in this setting.  
	Details for obtaining explicit values for the alternative using the \texttt{R} package {\bf \fontfamily{qcr} \selectfont MSPRT} are described in Section~S4.3 of the supplemental materials. The ${100(1-\alpha)}$th quantiles of a standard normal distribution and central $t$ distribution with $(N-1)$ degrees of freedom are denoted by $z_\alpha$ and $t_{\alpha;N-1}$, respectively, and $ \sigma $ denotes the known population standard deviation in a $z$ test, whereas $s_{n}$ refers to the sample standard deviation (with divisor $ (n-1) $) based on $n$ observations. 
	
	\end{tablenotes}
	\end{threeparttable}
	\end{table}

	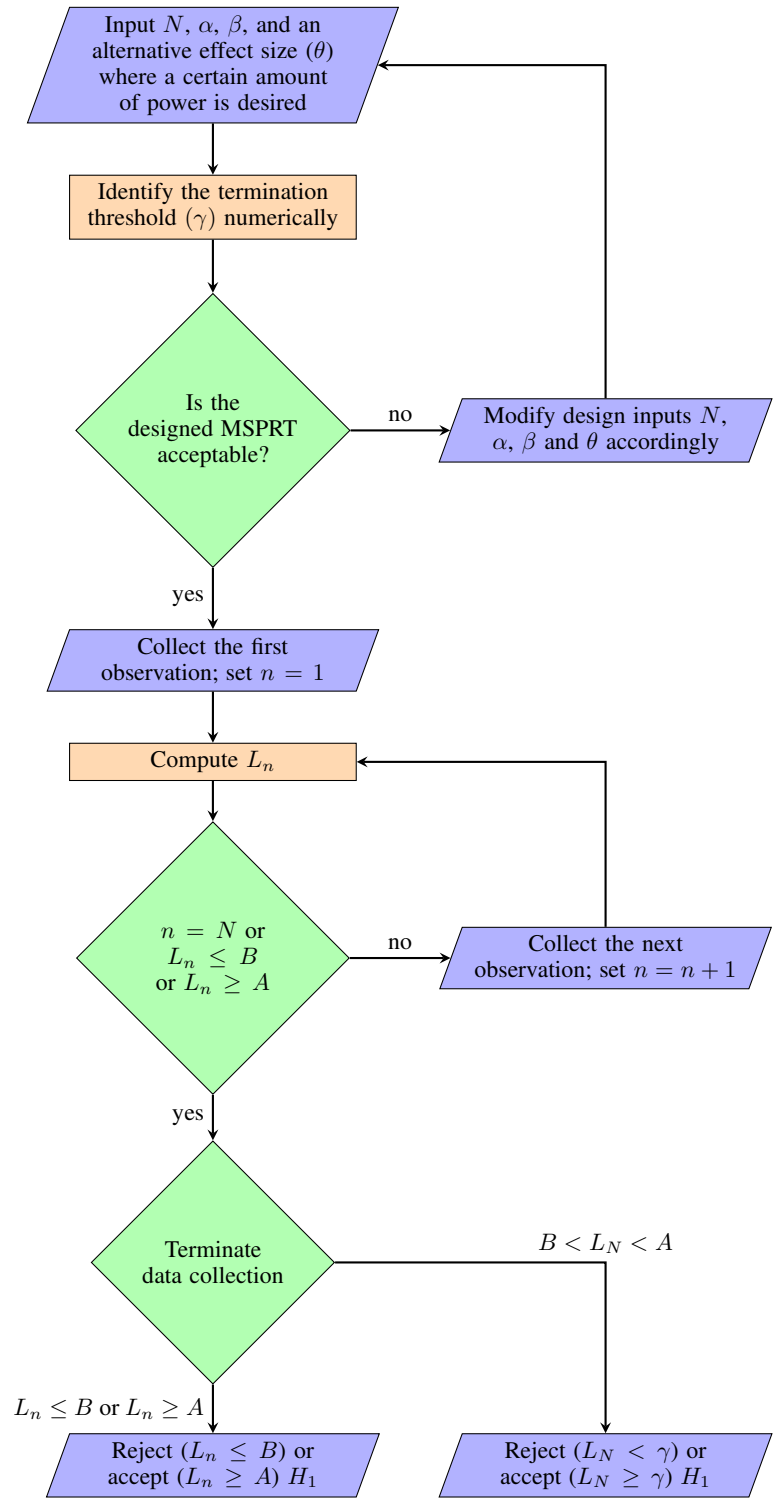
\begin{figure}[H]\hypertarget{flow}{}
		\centering
			\begin{tikzpicture}[scale=.9, every node/.style={scale=.9}]
			\node (in1) [io] {Input  $N$, $\alpha$, $\beta$, and an\\ alternative effect size ($\theta$) where a certain amount of power is desired};
			\node (pro1) [process, below of=in1, yshift=-1.1cm] { Identify the termination threshold $(\gamma)$ numerically};
            \node (dec1) [decision, below of=pro1, yshift=-2.3cm] {Is the\\ designed MSPRT acceptable?};
            \node (in4) [io, right of=dec1, xshift=4.8cm] {Modify design inputs $N$, $\alpha$, $\beta$ and $\theta$ accordingly};
			\node (in2) [io, below of=dec1, yshift=-2.4cm] {Collect the first\\ observation; set $ n=1 $};
			\node (pro2) [process, below of=in2, yshift=-.5cm] {Compute $L_n$};
			\node (dec2) [decision, below of=pro2, yshift=-1.9cm] {$n=N$ or $\Ln\leq B$ or $\Ln \geq A$};
			\node (in3) [io, right of=dec2, xshift=4.8cm] {Collect the next\\ observation; set $n=n+1$};
			\node (dec3) [decision, below of=dec2, yshift=-3.5cm] {Terminate data collection};
			\node (outa) [io, below of=dec3, yshift=-2cm] {Reject ($\Ln\leq B$) or accept ($\Ln\geq A$) $H_1$};
			\node (outb) [io, right of=outa, xshift=4.8cm] {Reject ($L_N < \gamma$) or accept ($L_N \geq \gamma$) $H_1$};
			\draw [arrow] (in1) -- (pro1);
			\draw [arrow] (pro1) -- (dec1);
			\draw [arrow] (dec1) -- node[anchor=south] {no} (in4);
			\draw [arrow] (in4) |- (in1);
			\draw [arrow] (dec1) -- node[anchor=east] {yes} (in2);
			\draw [arrow] (in2) -- (pro2);
			\draw [arrow] (pro2) -- (dec2);
			\draw [arrow] (dec2) -- node[anchor=south] {no} (in3);
			\draw [arrow] (in3) |- (pro2);
			\draw [arrow] (dec2) -- node[anchor=east] {yes} (dec3);
			\draw [arrow] (dec3) -- node[anchor=east] {$\Ln \leq B$ or $\Ln \geq A$ } (outa);
			\draw [arrow] (dec3) -| node[anchor=south] {$B< L_N <A$} (outb);
			\end{tikzpicture}
		\caption{A flow chart representing the MSPRT procedure.}
	\end{figure}
	
Figure \hyperlink{flow}{1} summarizes the process for conducting a MSPRT for a one-sided test of a normal mean or a population success probability.
\end{sloppypar}

\section{Implementation}\label{sec:Implementation}

\begin{sloppypar}
	Software to implement the MSPRT is available from the CRAN R software depository at \url{https://cran.r-project.org/web/packages/MSPRT/index.html} and on GitHub at \url{https://github.com/sandy-pramanik/MSPRT}. The software can be used to perform one-sample proportion tests, and one- and two-sample $z$ and $t$ tests. To design and implement a MSPRT, a user must provide a null hypothesis ($\theta_0$), a direction of the alternative hypothesis (right, left or two-sided), maximum available sample size ($N$), and pre-specified error probabilities ($\alpha$ and $\beta$). Given these design parameters, the {\bf \fontfamily{qcr} \selectfont R} package {\bf \fontfamily{qcr} \selectfont MSPRT} provides test results based on sequential entry of outcome data. Detailed illustrations are provided in Section~S4 of the supplemental materials.
\end{sloppypar}

\section{Simulation Studies}\label{sec:Simulation}

This section analyzes the performance of the MSPRT through simulation studies. For simplicity, we first investigate the performance in one-sample tests for a binomial proportion, $z$ tests, and $t$ tests. Next, we compare the performance of the MSPRT with group sequential (GS) designs. The extension of MSPRT designs to two-sample $z$ and $t$ tests is immediate. We also compare the performance with Sequential Bayes Factors (SBF) \citep{schon17}. Finally, we discuss the potential benefit that is offered by MSPRT designs when we decrease the $p$-value threshold for declaring statistical significance from $5\%$ to $0.5\%$. Throughout the section, $10^6$ replications were used to summarize the performance of the MSPRT.

\subsection{Performance in one-sample tests}\label{sec:Performance in one-sample tests}

\begin{sloppypar}
    We examine one-sample tests for a binomial proportion, $z$ tests, and $t$ tests of size $\alpha=0.05$ and $\alpha= 0.005 $. For simplicity, we examine one-sided tests with alternative hypotheses of the form $H_1: \theta>\theta_0$. We also assume that the targeted power of the test is 80\% (i.e., $\beta=0.2$).  Two-sided tests, tests of alternative hypotheses of the form $H_1: \theta<\theta_0$, and tests with different Type I or Type II error probabilities are handled similarly. We compare the resulting MSPRT's to standard fixed design tests having the same $\alpha$ level, sample size $N$ and Type II error probability $\beta=0.2$. Given $N$ and $\alpha$ for 
	fixed design tests, we define $ \theta_a$, the fixed design alternative, as the alternative parameter value that provides the specified $\beta$.
	
	
	We now describe the simulation settings used for analyzing the operating characteristics and ASN's of the MSPRT. 	
    
    For one-sample $z$ tests, observations are assumed to be independent and identically distributed random samples from a normal distribution with unknown mean $\theta$ and known variance $\sigma_0^2$. To study the performance of the MSPRT under the null hypothesis, we generate observations from a $N(\theta_0, \sigma^2)$ distribution. For performance under $H_1$, we use a $N(\theta_a , \sigma_0^2)$ distribution with the fixed design alternative, $\theta_a$, defined in the previous paragraph. 
    We note that it is possible to numerically compute the operating characteristics and ASN of the MSPRT prior to the onset of an experiment.  The simulation setup for the proportion test proceeds exactly as above where we simulate the data independently from a Bernoulli$(\theta)$ distribution. For our simulations, we use $\theta_0 = 0$ for the $z$ test and $\theta_0 = 0.5$ for the proportion test (Section~S4.2 in the supplement provides implementation details).  For $t$-tests, $\theta_a$ is interpreted as the standardized effect size $\theta/\sigma$. 
	\begin{figure}[h]
	    \centering
		\includegraphics[width=.8\linewidth]{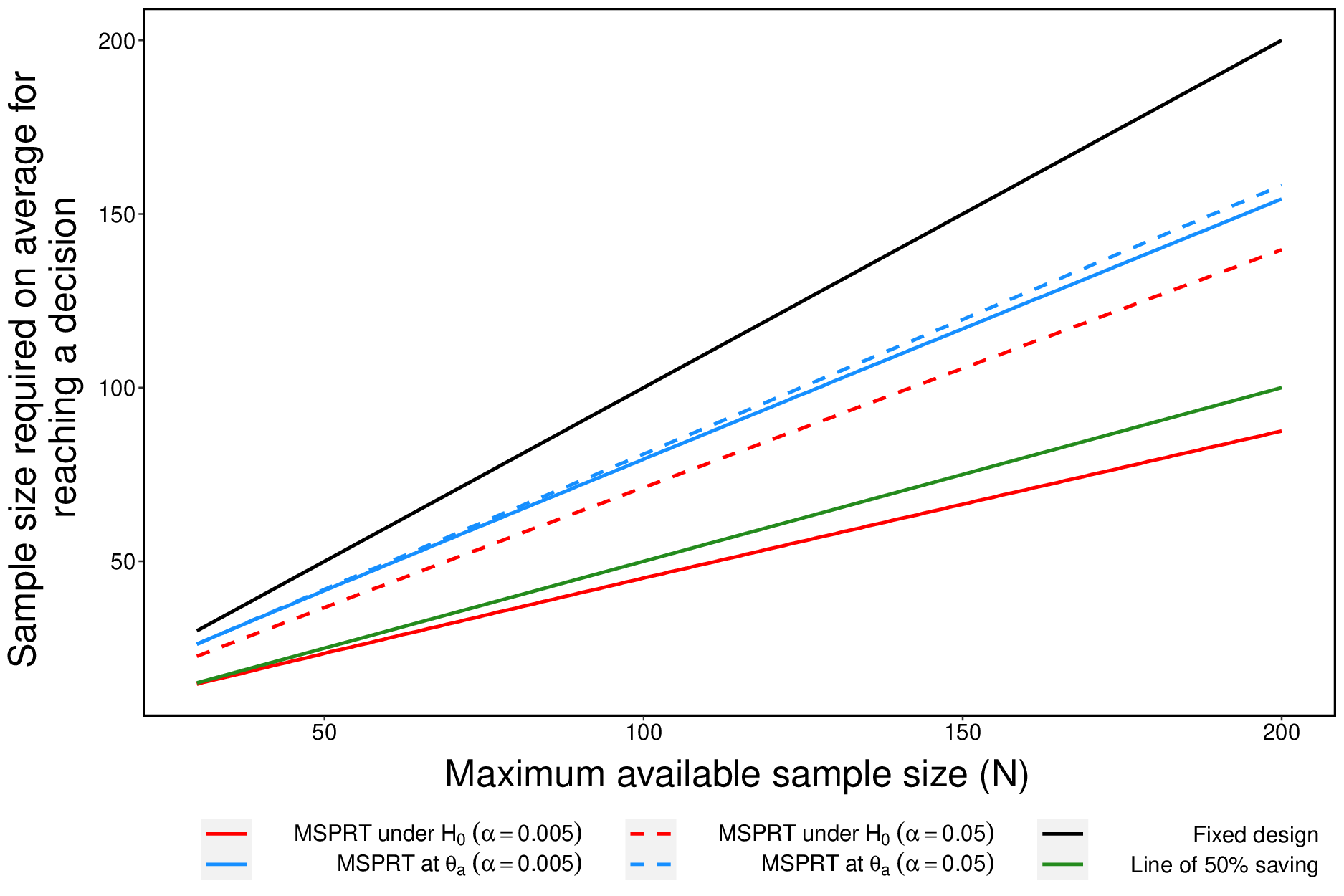}
		\caption{{\bf One-sample $t$ test that a population mean is 0.} Hypothesis test of $ H_0 : \theta = 0$ vs. $ H_1 : \theta > 0 $.  The population standard deviation is assumed to be unknown. Each curve in the plot represents the average number of samples, out of the maximum sample size ($N$), used before the MSPRT terminates in favor of the null or alternative hypothesis.  
			The operating characteristics under the alternative are evaluated at the corresponding fixed design point alternatives. 
		}\label{fig:onet}
	\end{figure}
	
	Figure~\ref{fig:onet} illustrates the performance of the MSPRT for a one-sample $t$ test of $H_0: \theta = 0$ versus $H_1:\theta > 0$. 
	This plot provides the average proportion of the $N$ samples required by a fixed design test for the MSPRT to achieve nearly equivalent Type I and Type II error probabilities.  Type I error probabilities are exactly maintained at targeted levels.  Type II error probabilities for the MSPRT's slightly exceed the targeted value of 0.2, but never exceed 0.22.
	
	The plot provided in Figure~\ref{fig:onet} for a one-sided $t$ test is nearly indistinguishable from the corresponding plots obtained for one-sample $z$ tests and tests of a binomial proportion; these plots are provided in the supplemental materials.  
	
	Two features of these plots are noteworthy.  First, for Type I error probabilities of $\alpha=0.005$, the average sample size required by the MSPRT is less than $ 50 \% $ of the sample size required by the fixed design test when the null hypothesis is true. This finding holds for all three tests. 
	Second, under the alternative hypothesis, the average sample size required for the MSPRT is typically about 80\% of the sample size required for the corresponding fixed design test.
	
	\textcolor{black}{To provide a theoretical context for these findings, we note that \cite{sieg1985} provides approximate formulae for power and the average sample number function for truncated SPRTs for large $ N $ (corollaries 3.45--3.47, page 55--57, Section 6 of Chapter III in \cite{sieg1985}). In order to derive these results, a Brownian motion process was used to approximate the operating characteristics and ASN of truncated SPRTs. 
	For $ \alpha = 0.005 $, the approximations predict that the average sample size required by the MSPRT under the null and the alternative hypothesis is approximately 40\% and 70\% of the fixed design sample size $ N $, respectively, and the Type II error probability at the fixed design alternative is about 23\%. These values match our empirical findings, but are based on approximating the discrete time scale of the MSPRT by the continuous time scale of Brownian motion. They also rely on an assumption that the test statistics are approximately normally distributed. As noted in \cite{sieg1985}, Brownian motion nonetheless furnishes a convenient way of analyzing properties of SPRTs while avoiding intractable probability calculations.}

	\begin{figure}[h!]
		\centering
		\includegraphics[width=.8\linewidth]{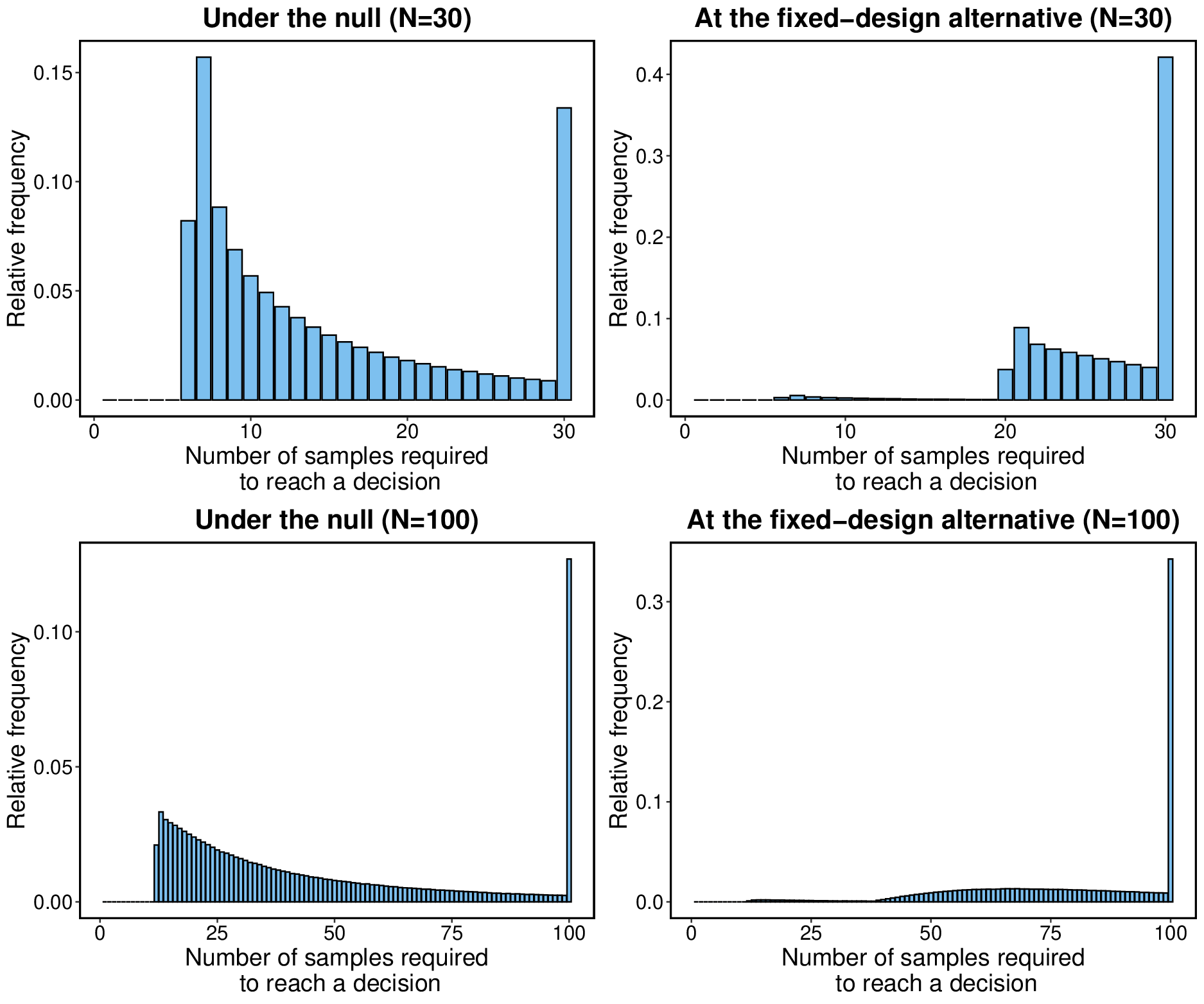}
		\caption{{\bf One-sample $t$ test that a population mean is 0.} Hypothesis test of $ H_0 : \theta = 0$ vs. $ H_1 : \theta > 0 $ \textcolor{black}{at $\alpha=0.005$ and $\beta=0.2$}.  The population standard deviation is assumed to be unknown. The barplots represent the distribution of sample size required by the MSPRT for reaching a decision under $H_0$ and at the corresponding fixed design alternative $\theta_a$. The fixed design alternatives, which provide 20$\%$ Type II error probability, are approximately 0.66 for $N=30$ and 0.35 for $N=100$.}\label{fig:stopping_time}
	\end{figure}
	Figure~\ref{fig:stopping_time} presents the distribution of the number of samples required by the MSPRT to reach a decision in a one-sample $t$ test. As in Figure~\ref{fig:onet}, the performance of the MSPRT is compared to the fixed design test having Type I error probability $0.005$ and Type II error probability $0.20$.  The top panel is based on a maximum sample size $N=30$, and the bottom panel on $N=100$. 
	From these figures, we see that under $H_0$ the MSPRT reaches a decision before the maximum sample size is accumulated in about $85\%$ of tests. This proportion slightly increases when $N$ is 100. Under the fixed design alternative, the MSPRT terminates in about $60\%$ of tests before the maximum sample size is reached when $N=30$, and in about $65\%$ of tests when $N=100$.  Though not displayed, similar results are obtained for one-sample proportion and $z$ tests.

	\begin{figure}[h!]
		\centering
		\includegraphics[width=.8\linewidth]{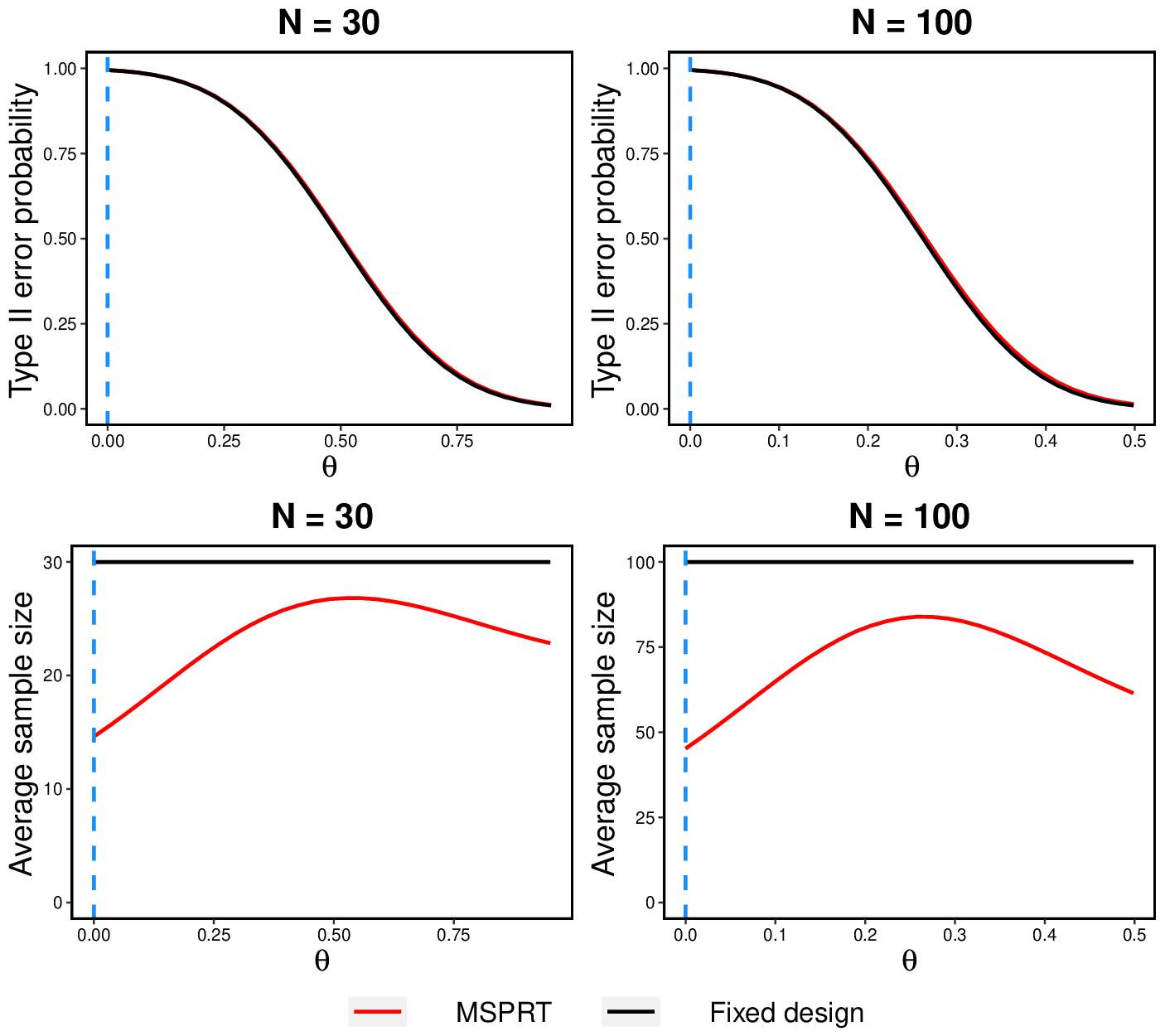}
		\caption{{\bf One-sample $t$ test that a population mean is 0.} Hypothesis test of $ H_0 : \theta = 0$ vs. $ H_1 : \theta > 0 $ at $\alpha=0.005$ and $\beta=0.2$.  The population standard deviation is assumed to be unknown. The above plots compare the Type II error probability and the average sample size of the MSPRT and the fixed design tests for a varied range of alternative effect sizes. The fixed design alternatives, which provide 20$\%$ Type II error probability, are approximately 0.66 for $N=30$ and 0.35 for $N=100$.
		}\label{fig:OCandASN}
	\end{figure}
	
	We also conducted a simulation study to analyze performance of the MSPRT at effect sizes other than the null and fixed-design alternatives $\theta_a$. Specifically, we considered the  right-sided one-sample $t$ test with $\alpha=0.005$ and $\beta=0.2$. We again set $N=30$  (left panel in Figure~\ref{fig:OCandASN}) and $N=100$ (right panel). To analyze the operating characteristics and ASN of resulting tests, we generated data using effect sizes that corresponded to fixed design tests having Type II error probabilities $0.05, 0.1, \cdots, 0.9$. The effect sizes range from 0.25 to 0.82 for $N=30$, and 0.13 to 0.43 for $N=100$. Figure~\ref{fig:OCandASN} presents the operating characteristics and ASN at these effect sizes. From the top panel we see that the Type II error probabilities of the MSPRT at these effect sizes are almost identical to the corresponding fixed design test (the red line almost coincides with the black line). Thus, the MSPRT achieves almost identical power to the fixed design test at a lower cost. The bottom panel in the same figure displays the ASN of the corresponding MSPRT's at the same effect sizes. The ASN's in this plot are about 70--80$\%$ of $N$ when the Type II error probability of the fixed design tests is 0.05 or smaller. 
	As the effect size decreases and gets closer to the null value, the ASN's increase until they reach a maximum of about 85--90$\%$ of $N$ for Type II error probabilities near 0.5. The ASN then  decreases to approximately 40--45$\%$ of $N$ near the null value. The performance of the MSPRT for other tests is  similiar to that depicted in Figure~\ref{fig:OCandASN}.

	\subsection{Comparison of MSPRT and GS designs}\label{sec:Comparison of the MSPRT with Group Sequential (GS) designs}
	
	We next compared the MSPRT to GS designs using the {\bf \fontfamily{qcr} \selectfont R} software package {\bf \fontfamily{qcr} \selectfont gsDesign} \citep{anderson2014gsdesign}. We used the default Hwang--Shih--DeCani error spending function as the sequential stopping criterion. For illustration, we assumed the design had a total of 5 groups/stages (including interim and final analysis) with equal number of subjects entered at each stage. As before, we varied $N$ from 30 to 200, considering two choices of the Type I error probability (0.05 and 0.005), and set the Type II error probability at 0.2.
	
	The {\bf \fontfamily{qcr} \selectfont gsDesign} function obtains the critical boundaries of the GS design by assuming a standard normal test statistic.  We therefore conducted a right-sided one-sample $z$ test of the form $H_0 : \theta=0$ vs  $H_0 : \theta>0$ with a known variance of 1. For a comparison under $H_1$, we focused at the fixed design alternative $(\theta_a)$ corresponding to design parameters $N$, $\alpha$ and $\beta$. After $\alpha$, $\beta$ and $\theta_a$ are specified in the {\bf \fontfamily{qcr} \selectfont gsDesign}, the software designs a test by exactly spending $\alpha$ and $\beta$ (at $\theta_a$) but with a slightly larger maximum sample size (than $N$). To make a fair comparison, we designed the MSPRT using this larger sample size as the maximum available sample size. We also adjusted the design parameters $\beta$ and $\gamma$ so that the designed MSPRT has approximately $1- \beta$ power (within $1\%$) at $\theta_a$.
	\begin{figure}[h!]
	    \centering
		\includegraphics[width=.8\linewidth]{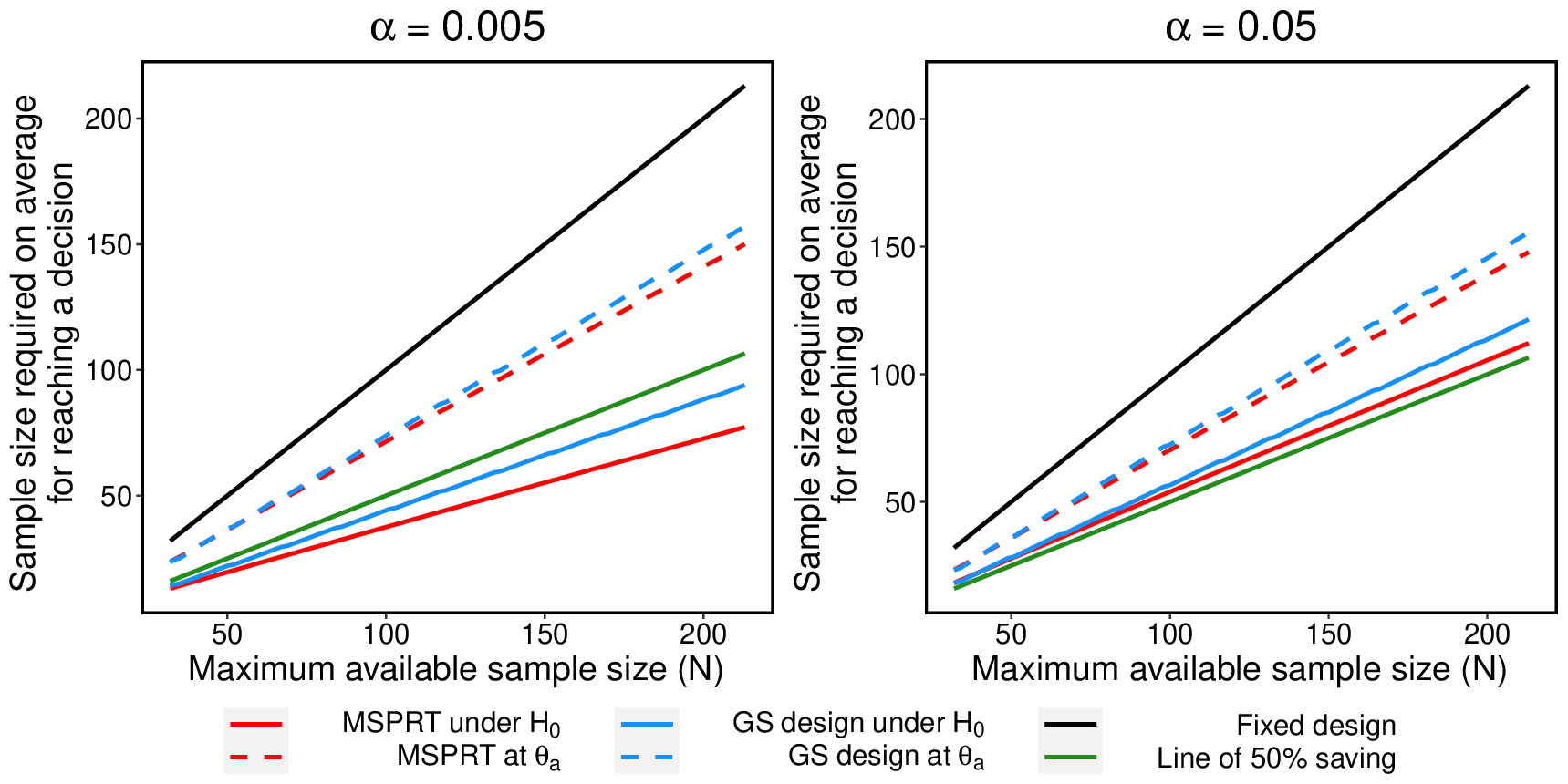}
		\caption{{\bf One-sample $z$ test that the population mean is 0.} Hypothesis test of $ H_0 : \theta = 0$ vs. $ H_1 : \theta > 0 $. Each curve in the plot represents the average number of samples, out of the maximum sample size ($N$), used before the MSPRT or the GS design terminates in favor of the null or alternative hypothesis.  
		}\label{fig:gsdesign comparison}
	\end{figure}
	
	In Figure~\ref{fig:gsdesign comparison} we compare the average sample size used in the MSPRT and GS tests. For a varied range of $N$, the MSPRT achieves a uniformly smaller ASN than the GS design. Their performances are quite similar under both $H_0$ and $\theta_a$ when $\alpha=0.05$, and at $\theta_a$ when $\alpha=0.005$. A more visible difference can be seen under $H_0$ when $\alpha=0.005$. At the higher significance level, the GS design uses about $44\%$ of the maximum available sample size. The MSPRT on an average uses about 3--8$\%$ fewer samples for the same Type I and Type II error probabilities. The difference in ASN becomes larger as the maximum available sample size increases.

	{\color{black}
	\subsection{Performance comparison between MSPRT and SBF in two-sample $t$ tests}\label{sec:Comparison with the SBF}
	
	In this section we compare the performance of the MSPRT to the sequential Bayes factor (SBF; \cite{schon17}).
	 At each step of a sequential analysis, a SBF computes the Bayes factor under a Cauchy prior on the standardized effect size. The stopping boundaries 
	 are based on verbal labels for grades of evidence \citep{Kass&Raftery1995}. We note that SBF tests, like the SPRT, do not fix maximum sample sizes in advance.
	 
	 We make this comparison for two-sided two-sample $t$ tests because of their widespread application.  Let $ \theta_1 $ and $ \theta_2 $ be the population means of two groups of subjects. Under the assumption that the observations from the underlying populations are normally distributed and their common variance is unknown, a two-sided two-sample $t$ test compares the hypothesis $ H_0: \theta_1 - \theta_2 =0 $ against the alternative hypothesis $H_1: \theta_1 - \theta_2 \neq 0$. To conduct this two-sided test with Type I error probability $\alpha$, a MSPRT simultaneously performs two separate one-sided tests, each with Type I error probability $\alpha/2$. At each sequential step it $(i)$ rejects $H_1$ if both the tests reject $H_1$, $(ii)$ rejects $H_0$ if either test rejects $H_0$, or $(iii)$ continues sampling.
	 
	 To simplify exposition, we assume the maximum number of subjects available in both groups is equal and is denoted by $N$, and that sequential testing is performed so that one pair of subjects from each group are measured simultaneously.  The total sample size for the experiment is thus $2N$.

\begin{figure}[h]
	    \centering
		\includegraphics[width=.8\linewidth]{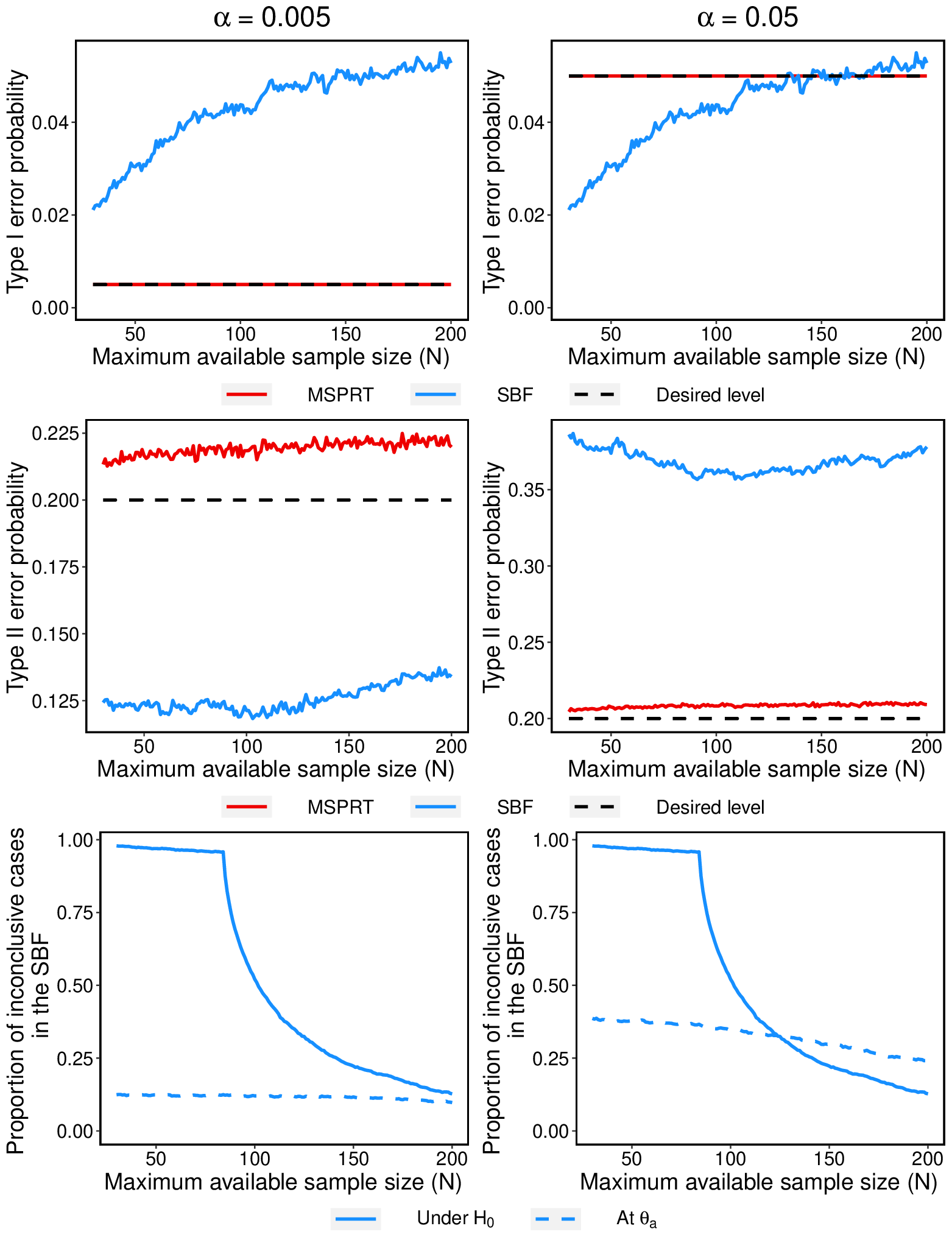}
		\caption{\color{black}{\bf Comparison of error probabilities for SBF and MSPRT tests.} Two choices for the targeted Type I error probabilities of 0.005 (left column) and 0.05 (right column) for the MSPRT are considered. For both the tests we varied the maximum available sample size $(N)$ and compared the Type I (first row) and the Type II (second row) error probabilities achieved. The final column displays the proportion of inconclusive cases at the maximum sample size for the SBF.
		}\label{fig:sbfdesign error comparison}
	\end{figure}

\begin{figure}[h]
	    \centering
		\includegraphics[width=.8\linewidth]{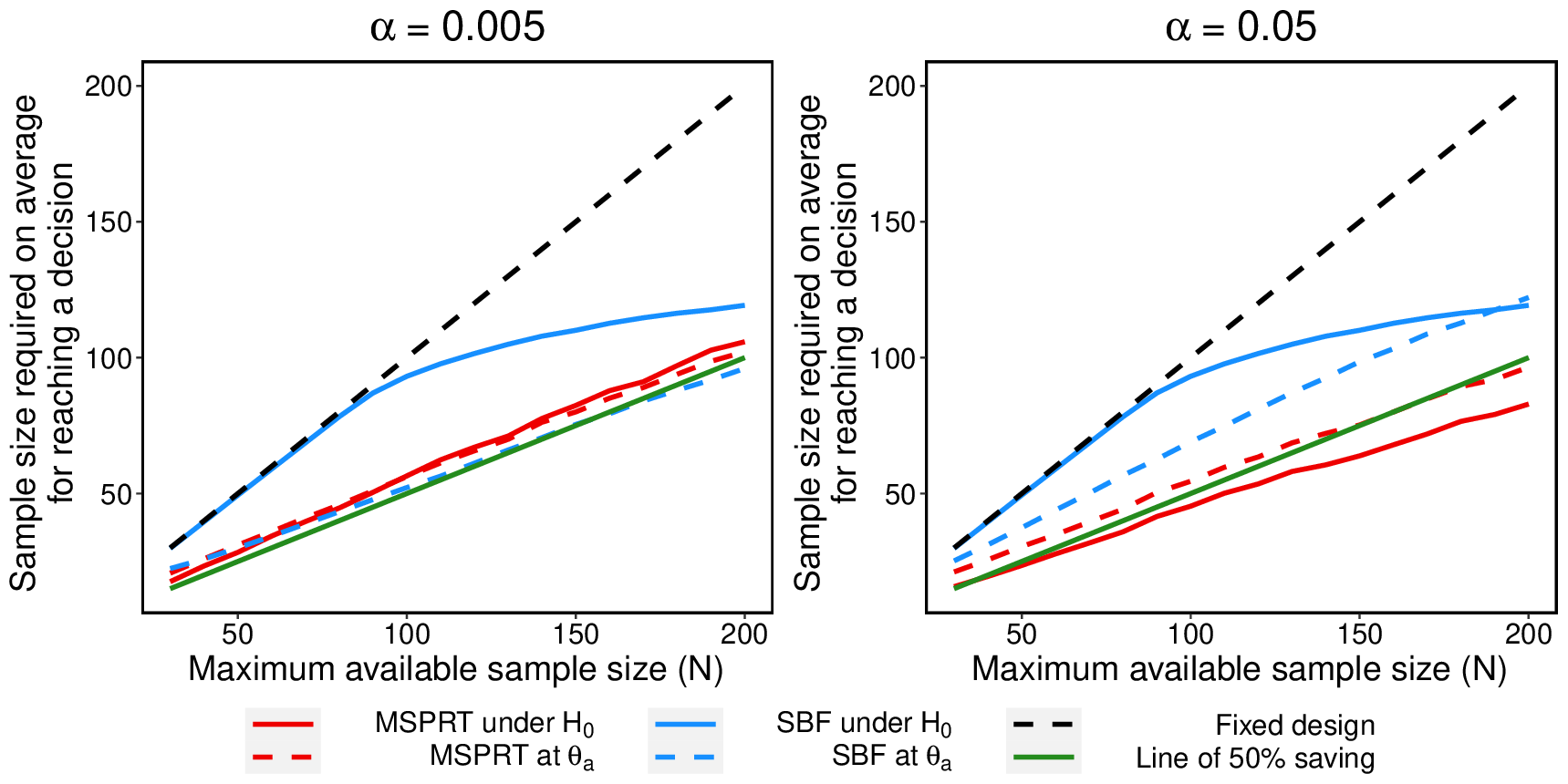}
		\caption{\color{black} {\bf Comparison of ASN for MSPRT and SBF.} This plot displays the proportion of the maximum sample size under various assumptions on null and alternative hypotheses for the MSPRT and SBF tests.
		}\label{fig:sbfdesign asn comparison}
	\end{figure}
	
	Figure~\ref{fig:sbfdesign error comparison} presents a comparison between the two sequential procedures for testing $ H_0 : \theta_1 - \theta_2 = 0$ against $ H_1 : \theta_1 - \theta_2 \neq 0 $. The performance under $H_1$ is examined at the corresponding fixed-design alternatives $(\theta_a)$. Figure~\ref{fig:sbfdesign error comparison} presents results for the right-sided alternative $\theta_a$, the results for the left-sided alternative being similar.  
	To implement the SBF test, we followed the recommendations of \cite{schon17} and set the Cauchy scale parameter $r$ equal to 0.707. The null and alternative boundaries for the Bayes factor were fixed at $1/6$ and $6$, respectively, and the minimum sample size was set to 20 in each group. We assumed that a maximum of $N$ samples were available for each group, and the sample sizes in the two groups were equal. If the SBF test did not reach a decision after accruing all subjects (i.e., $1/6 <  \mbox{SBF} < 6$), it was assumed that the test failed to reject the null hypothesis. Such outcomes thus decrease the Type I error of the SBF tests. 
	
	Because the goals and philosophy underlying the SBF and MSPRT are different, choosing the design parameters for the MSPRT to make a comparison to the SBF is difficult. For this reason, we choose two default settings for the MSPRT corresponding to  $\alpha=0.05$ and $\alpha=0.005$, holding $\beta=0.2$.  In all comparisons,  we assumed that pairs of observations were collected sequentially until each test terminated (possibly at the maximum sample size $2N$). 
	We emphasize that the SBF is not intended to control either Type I or Type II error probabilities, so achieving these rates should not be regarded as a basis of comparison.
	
	

    Figure~6 displays results for this comparison. The first row shows that that the MSPRT achieves its targeted Type I error probability for both tests.  The Type I error achieved by the SBF is identical in both plots since the design parameters of that test did not change.
    
    The second row of plots in Figure~6 displays the Type II error of each test, at the alternative targeted by the MSPRT.  When $\alpha=0.005$, the Type II error probability of the MSPRT is higher than the SBF, while it is lower for $\alpha=0.05$.  The Type II error probability for the SBF changes between plots because the alternative being tested has changed. 
    The final row of this plot indicates the proportion of tests that were inconclusive at the maximum sample size for the SBF test.

    Because the SBF does not control error probabilities at pre-assigned values, additional care is needed to compare the ASN needed for each test. To make such a comparison, we therefore implemented the following procedure.  At each $N$, we determined the (positive) value of the alternative hypothesis that provided 80\% power in a fixed-design, two-sided $t$ test with Type I error probability of either 0.05 or 0.005, against a null hypothesis of 0.  Through simulation, we then determined the Type I and Type II error probabilities of the SBF (using the truncation rule described above).  We then constructed the MSPRT with the same (within 1\% numerical error) Type I and Type II error probabilities.  
    This procedure allowed us to compare the average sample sizes of the two testing procedures with approximately similar error probabilities. The resulting comparison of the average sample sizes is presented in Figure~7. }
    
    \color{black}{For $\alpha=0.005$, Figure~7 suggests that both tests require approximately the same ASN when the alternative hypothesis is true. However, the MSPRT is substantially more efficient when the null hypothesis is true.   In both plots, the solid black curve represents the ASN for the SBF when the null hypothesis is true, and this curve falls well above the corresponding solid red curve representing the ASN for the MSPRT. 
    }
	
	\color{black}{
	The SBF test's use of a median-zero Cauchy prior to define the alternative hypothesis provides a partial explanation of these findings. This prior assigns significant mass around 0, the hypothesized effect size under the null. The Cauchy prior is a particular example of a local prior, and it is known that the evidence in favor of a true null hypothesis accumulates much slower than it does under a true alternative hypothesis when local priors are used \citep{Johnson&Rossell2010}. To fix this asymmetric accumulation of evidence, Johnson and Rossell proposed non-local priors on effect sizes under alternative hypotheses which assign zero prior density to the null value. Since the UMPBT alternatives place all their mass at non-null effect sizes, they are non-local alternative priors and can thus be expected to accumulate evidence more rapidly in favor of true null hypotheses.  
	
	For tests based on fixed sample sizes, we note that UMPBTs (when they exist) are, by definition, the tests that provide the highest probability of exceeding a specified Bayes factor threshold.}

	\subsection{Higher significance with similar sample sizes}\label{sec:Offsetting the cost of decreasing alpha}
	
	We next examine the potential benefit that the MSPRT offers in offsetting the increase in the sample size that would be required if the bar for declaring a result ``statistically significant'' were moved from $p<0.05$ to $p<0.005$. Specifically, we compare the sample size required in standard fixed design tests at the 5\% level to the average sample size required by the MSPRT at the 0.5\% level.
	
	If the null hypothesis is true, this comparison is straightforward.  If not, care must be taken to make sure that the same alternative hypotheses are compared at both levels of significance in the fixed design and MSPRT design scenarios.  To make this comparison, we determine the $\theta^*$ that achieves the targeted Type II error probability in a fixed design test of size $\alpha=0.05$.  For that $\theta^*$, we next determine the $N^*$ needed to achieve the same Type II error probability in a fixed design test of size $\alpha=0.005$.   We then set that $N^*$ as the maximum sample size for the MSPRT.
	
	Because the average sample size used in the MSPRT depends on whether the null or alternative hypothesis is true, and because we are interested in the long-run effect of implementing the MSPRT over many experiments, it is useful to examine the effect on the total sample size as the proportion of true null hypotheses is varied.  Recent research suggests that this proportion is likely to be in the range 0.80--0.95 \citep{dreber15,johnson17}.

	\begin{figure}[h]
		\centering
		\includegraphics[width=.8\linewidth]{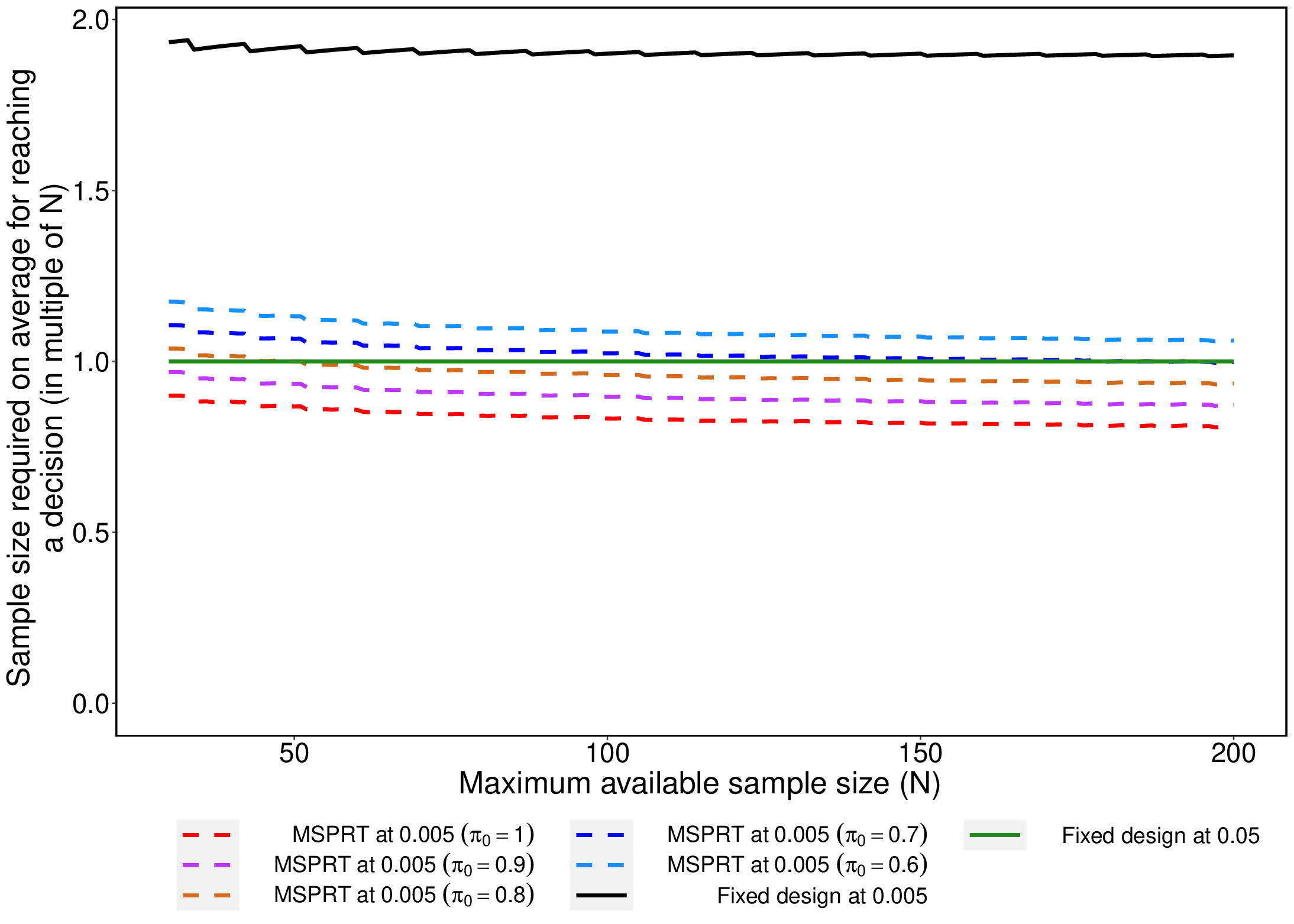}
		\caption{{\bf One-sample $t$ test that a population mean is 0.} Curves in this plot represent the average multiple of the sample size in a fixed design test of size $\alpha=0.05$ required to perform the MSPRT of size $\alpha=0.005$ of approximately the same power. Average sample sizes are dependent on the proportion of tested null hypotheses that are true. The MSPRT maintains a Type I error probability of 0.005, and its power at $ \theta^* $ always exceeds 0.77 for the indicated proportion of $N^*$ (the sample size of the corresponding fixed design test).}\label{fig:onet005vs05}
	\end{figure}

	In the case of a one-sample $t$ test, Figure \ref{fig:onet005vs05} displays the average multiple of the fixed 5\% test's sample size $N$ that is required to perform the MSPRT with size 0.5\% as the proportion of tested null hypotheses $\pi_0$ is decreased from 1 (the dashed red line at the bottom) to 0.6 (the light black line).  Also displayed is the multiple of $N$ that is required to achieve a Type I error probability of 0.005 in a fixed design test (the solid black line at the top).  The latter multiple tends to fall between 1.89 and 2.14. Similar plots are obtained for one-sample $z$ tests and tests of a binomial proportion; these plots are provided in the supplemental materials.
	
	The key finding from Figure \ref{fig:onet005vs05} is that MSPRTs for $ \alpha = 0.005 $ require, on average, essentially the same sample sizes that are required to conduct one-sided, fixed design tests for $ \alpha = 0.05 $ for tests designed to have Type II error probabilities of 0.2. We emphasize that such gains may not be achieved at tests implemented with more stringent Type II error probabilities or in two-sided test, and it is important to study the operating characteristics of any particular design before its implementation. In the case of one-sided $z$, $t$ and proportion tests, however, this finding holds because $ N^* $ is roughly two times that of $ N $, but at $ \alpha = 0.005 $ the MSPRT saves more than $ 50\% $ of the maximum available sample size when the null hypothesis is true and the test is powered to achieve a Type II error of $0.2$. For such tests, ``raising the significance bar" from 0.05 to 0.005 could be accomplished without significantly increasing sample sizes if MSPRTs were used in place of fixed design tests.
\end{sloppypar}

{\color{black}

\bigskip
\section{An Application to the retrospective gambler’s fallacy study}\label{sec:realdata}

\begin{sloppypar}
	In this section we illustrate the use of the MSPRT to the replication data of the retrospective gambler’s fallacy study, one of many studies available from the first ``Many Labs'' project \citep{klein2014}. The data is openly accessible from the Open Science Framework (\url{https://osf.io/wx7ck/}). In the original study, \cite{oppenheimer2009} investigated the influence of observing a rare, independent, chance event on individuals' perception of preceding events. For the experiment, the participants imagined that they saw someone rolling dice in a casino and then witnessed one of the following three outcomes (or conditions). In one condition, the participants imagined that they observed three dice being rolled and all came up ``6'' (the ``three6 condition''). In the second condition, two dice came up ``6'' and one was ``3'' (the ``two6 condition''). Finally, in a third condition, two dice were rolled and both came up ``6.''  All participants then estimated the number of times the dice were rolled before they observed the outcomes. The results from the study support a theoretical prediction that participants perceive unlikely outcomes to have arisen from longer sequences than more common outcomes. 
	
	In the Many Labs project, the same study was replicated with only the first and second conditions. In that study, there were a total of 5942 participants out of whom 2680 participants witnessed the three6 condition and 3262 witnessed the two6 condition. To keep the illustration simple we consider a sequential MSPRT with equal number of participants from each group. Thus, we randomly selected 2680 participants from the two6 group and the full set of 2680 responses from the three6 condition as our data for the sequential MSPRT.  Furthermore, following \cite{oppenheimer2009} and \cite{klein2014}, we took the square-root of the subjects' estimated number of dice rolls prior to their imagined outcome as the response variable.  (The square-root transformation of Poisson counts is approximately variance stabilizing).
	
	To test the hypothesis of a mean difference, we assumed that the underlying population of the transformed responses corresponding to the three6 and two6 conditions were independently and normally distributed with means $\theta_3$ and $\theta_2$ with an unknown common variance $\sigma^2$. We then applied a right-sided two-sample $t$-test of the form 
	
	\begin{equation}\label{eq3}
	H_0 : \theta_3 - \theta_2 = 0 \qquad \mbox{vs.} \qquad \theta_3 - \theta_2 > 0
	\end{equation}
	with the Type I and the Type II error probabilities constrained by $\alpha$ and $\beta$, respectively.
	
	Approximately 90 subjects, on average, were assigned to each group in the Many Labs project, so we arbitrarily set $N=90$ in this study.  We then varied  $\alpha = \{ 0.005, 0.05 \}$ and $\beta = \{ 0.05, 0.2 \}$ and examined the operating characteristics of the MSPRT by repeatedly sampling 90 subjects from the two treatment groups. 
	
	For each $(\alpha, \beta)$ combination, we designed the MSPRT using the {\bf \fontfamily{qcr} \selectfont design.MSPRT()} function in the {\bf \fontfamily{qcr} \selectfont R} package {\bf \fontfamily{qcr} \selectfont MSPRT}. For example, when $\alpha = 0.005$ and $\beta = 0.05$, the {\bf \fontfamily{qcr} \selectfont R} command to obtain the MSPRT design parameter is as follows:
	\begin{verbatim}
	# design the MSPRT
	>out = design.MSPRT(test.type = 'twoT',
	                    Type1.target = 0.005,
	                    Type2.target = 0.05,
	                    N1.max = 90, N2.max = 90)
	
	# display termination threshold
	>out$termination.threshold
	
	# display simulation estimate of Type I error 
	# probability obtained by the MSPRT
	>out$Type1.attained
	
	# display simulation estimate of Type II error 
	# probability obtained by the MSPRT at 
	# the fixed-design alternative
	>out$Type2.attained
	
	# display the ASN of the MSPRT under the null
	>out$EN$H0
	
	# display the ASN of the MSPRT at the fixed-design 
	# alterative
	>out$EN$H1
	\end{verbatim}
	
		We next applied the MSPRT to actual data sets by randomly selecting 90 (sequential) observations from the available 2680 observations in each treatment group.  For  $\alpha = 0.005$ and $\beta = 0.05$, applying the MSPRT to the first sequence of sampled outcomes led to rejection of the null hypothesis at the 0.005 level of significance after $60$ observations were observed from each group. The MSPRT was implemented using the {\bf \fontfamily{qcr} \selectfont implement.MSPRT()} function as follows:
			\begin{verbatim}
	>implement.MSPRT(obs1 = three6.resp.MSPRT, 
	                 obs2 = two6.resp.MSPRT,
	                 design.MSPRT.object = out)
	\end{verbatim}
	Here, \texttt{three6.resp.MSPRT} and \texttt{two6.resp.MSPRT} are numeric vectors containing the sequential observations of the three6 and two6 responses, and \texttt{out} is the object storing the MSPRT output. The {\bf \fontfamily{qcr} \selectfont implement.MSPRT()} command can be executed sequentially after responses are observed and the response variables have been updated. The Bayes factor obtained at the MSPRT alternative for this sequence of observations is displayed in Fig.~9.
	
	\begin{figure}
		\centering
		\includegraphics[width=.8\linewidth]{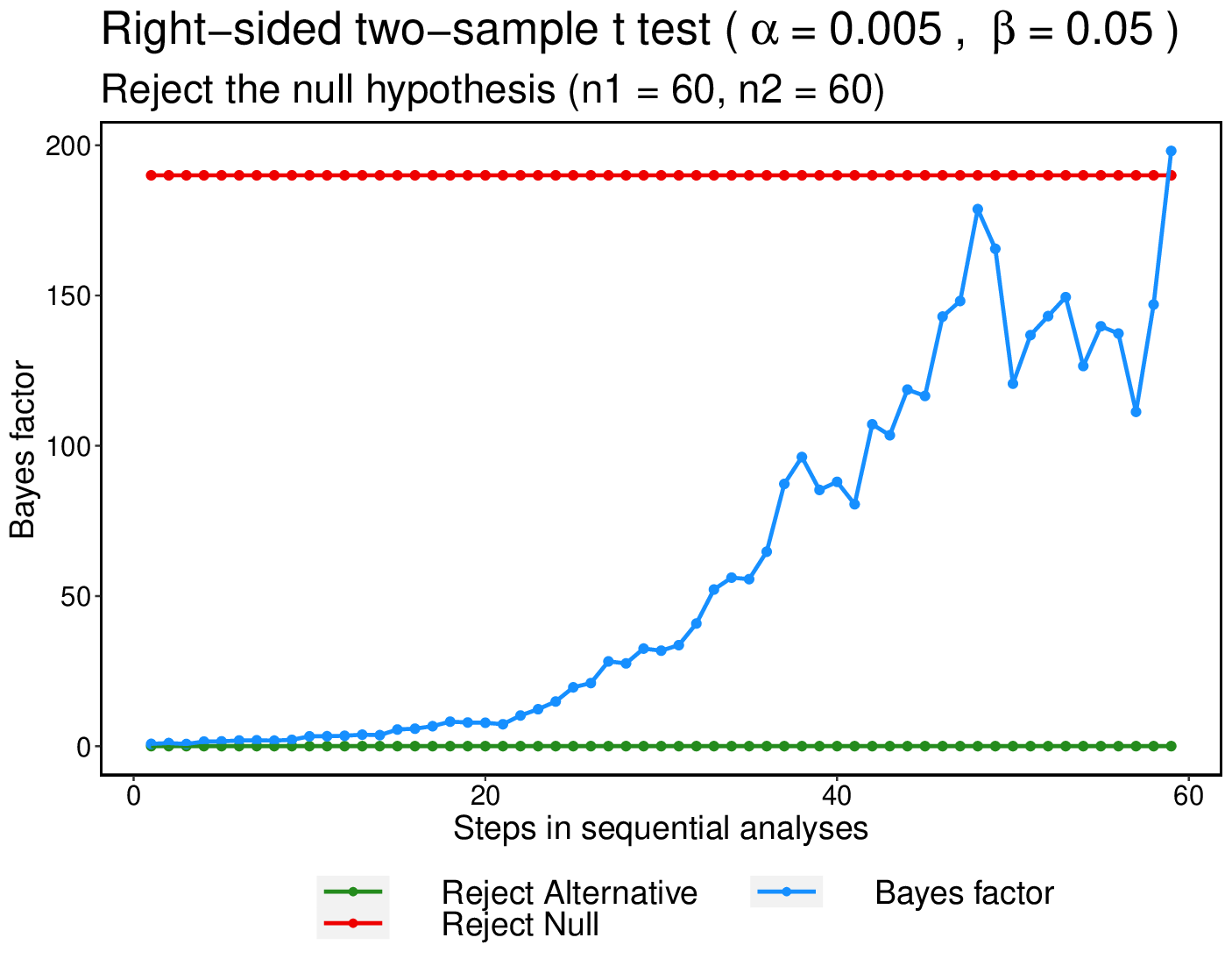}
		\caption{\color{black}Application of the MSPRT at $\alpha=0.005$ and $\beta=0.05$ to a specific simulated sequence of observations from each group available from the retrospective gambler’s fallacy study.}\label{fig:gambling}
	\end{figure}

	We can also find the operating characteristics of the MSPRT at specified effect sizes using the {\bf \fontfamily{qcr} \selectfont OCandASN.MSPRT()} function. For example, the {\bf \fontfamily{qcr} \selectfont R} commands to calculate the operating characteristics of this MSPRT, at the estimated standardized effect of 0.69 cited in \citep{oppenheimer2009} are as follows: 
	\begin{verbatim}
	# obtain the OC at theta = 0.69
	>oc.out = OCandASN.MSPRT(theta = 0.69,
	                         design.MSPRT.object = out)
	
	# display simulation estimate of Type II error probability 
	# at theta = 0.69
	>oc.out$acceptH0.prob
	
	# display ASN from Group-1 at theta = 0.69
	>oc.out$EN1
	
	\end{verbatim}
	The output from these commands, \texttt{oc.out}, is a data frame in which rows correspond to effect sizes, and columns refer to the probability of rejecting $H_1$ and the ASN's from Group 1 and Group 2 (in the case of equal sample sizes in both groups, these values are the same). For reference, these values are displayed in 
	Table~\ref{table: gambler MSPRT designs}.

	\begin{figure}
		\centering
		\includegraphics[width=.8\linewidth]{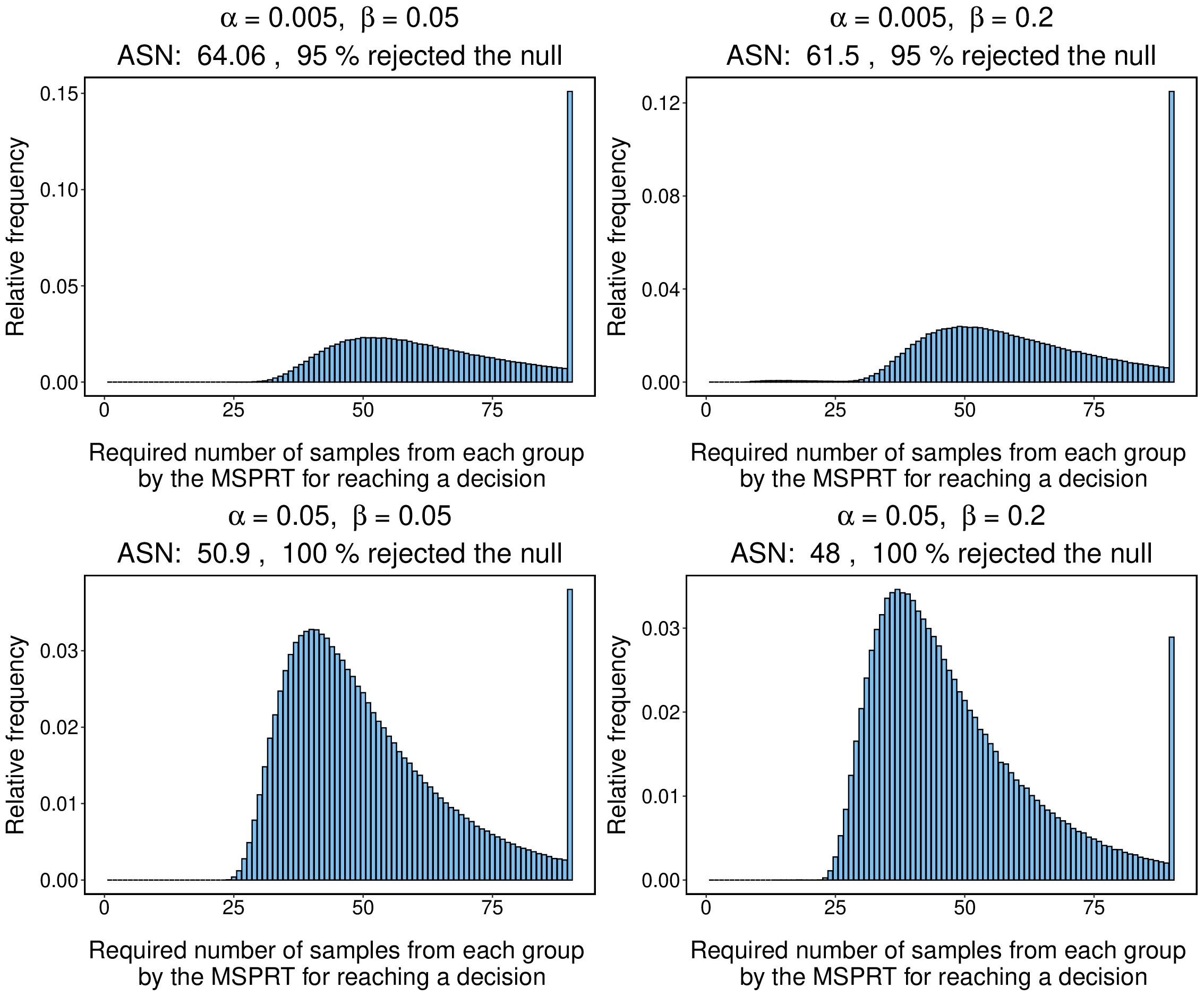}
		\caption{\color{black} Histogram of the required number of samples from each group (condition) by the MSPRT for reaching a decision in $10^6$ random permutations of the gambler’s fallacy study responses. }\label{fig:gambling stopping time}
	\end{figure}
	
	For each pair of $(\alpha, \beta)$, we also evaluated the operating characteristics of the MSPRT when applied to $10^6$ sampled sequences.   Specifically, we calculated $(a)$ the number of samples required on average by the MSPRT to reach a decision, and $(b)$ the proportion of times the MSPRT rejected the null hypothesis. These results are presented in Figure~\ref{fig:gambling stopping time}. 
\end{sloppypar}
}

	\begin{table}
		\centering
	\begin{threeparttable}
		\caption{\em Operating characteristics and ASN's of the designed MSPRT's for the retrospective gambler’s fallacy study}\label{table: gambler MSPRT designs}
		\begin{tabular}{c c c c c c c c}
				\midrule[1.3pt]
				\multirow{2}*{$\alpha$} & \multirow{2}*{$\beta$}	&   \multicolumn{2}{c}{ $\bigtriangleup = 0$} &   \multicolumn{2}{c}{ $\bigtriangleup  = \theta_a$} &   \multicolumn{2}{c}{ $\bigtriangleup = 0.69$} \\ [0.5ex]
				\cmidrule(lr){3-4} \cmidrule(lr){5-6} \cmidrule(lr){7-8}
				& 	& Type I    & $ \mathbb{E} (n)$	& Type II	& $ \mathbb{E} (n)$  & Type II	& $ \mathbb{E} (n)$\\ [0.5ex] \midrule
				
				\multirow{2}*{0.005}	&   0.05 & 0.005 &   63.46   &   0.0513   & 63.31   & 0.023 & 58.4	\\ \cmidrule (l){2-8}
				
				&   0.2 & 0.005 &   39.83  &   0.2129   &  69.78  & 0.029 & 56.06	\\ \midrule
				
				\multirow{2}*{0.05}	&   0.05 & 0.05 &   84.3   &   0.0504   & 63.74   & 0.001 & 46.77	\\ \cmidrule (l){2-8}
				
				&   0.2 & 0.05 &   63.28  &   0.204   &  71.52  & 0.002 & 44.13	\\
				\midrule[1.3pt]
		\end{tabular}
		\begin{tablenotes}
		\small
		\item{\em Note.} Type I and Type II indicates the corresponding error probabilities. $\bigtriangleup = (\theta_3-\theta_2)/\sigma $ denotes the standardized effect size. $ \mathbb{E} (n)$ denotes the ASN for each group at the corresponding effect size. Effect sizes at the null value $\bigtriangleup=0$, fixed-design alternative $\theta_a$ (i.e., the fixed $N$ design providing the specified $(\alpha,\beta)$), and the standardized effect size 0.69 estimated from the original study are provided.
	
	\end{tablenotes}
	\end{threeparttable}
	\end{table}

\section{Discussion}\label{sec:Discussion}

\begin{sloppypar}
	The costs of conducting experiments to test hypothesized effects is often related directly to the number of tested items or 
	participants.  When 
	the study data can be collected sequentially, the use of sequential testing procedures can dramatically reduce these costs.  When tests are designed to identify 
	hypothesized effects that do not exist (i.e., the null hypothesis is true), the use of the MSPRT can reduce the sample sizes to reach a decision.  \textcolor{black}{In $z$ and $t$ tests with type II error probabilities targeted at 20\%, the reduction in sample sizes can be as much as 20\% to 30\% in 5\% tests, and as much as 50\% in 0.5\% tests. }

	Much of this article has focused on one-sample $z$, $t$, and proportion tests.  Mathematically, two-sample $z$ and $t$ tests are similar to one-sample tests, and so our findings extend to two-sample $z$ and $t$ tests.
	Table S1 of \cite{johnson13_1} provides a list of the UMPBT alternatives and the likelihood ratios (or Bayes factors) for two-sample $z$ and $t$ tests. Section S4 of the supplemental materials provides a user guide for two-sample tests.
	
	A potential drawback in the implementation of MSPRTs is the firm requirement to specify the outcome variable and test statistic prior to the start of the experiment. Of course, in principle the same requirement applies to fixed design experiments, but failure to ensure that these quantities are clearly identified a priori could lead to additional opportunities for $p$ hacking and other unethical practices in sequential designs.   For instance, researchers might apply MSPRTs to several outcome variables simultaneously, which would negatively affect the control of Type I errors.  In addition, the conduct of MSPRTs requires that investigators perform statistical analyses after the acquisition of each participant's data, which in some settings may not be feasible. However, for studies in which a high threshold for significance is desired, this technique may offer researchers a method of testing hypotheses while maintaining required sample sizes at a manageable level.

\end{sloppypar}

\section{Supplementary Materials}\label{sec:Supplementary Materials}

\begin{sloppypar}
	Supplementary materials, which are available online, contain a detailed discussion of the UMPBT alternative and a comprehensive user guide for the {\bf \fontfamily{qcr} \selectfont MSPRT} package. Section S2 highlights the general MSPRT for testing a simple null against a compositive alternative hypothesis. In Section S3, the UMPBT alternatives are discussed in detail for one-sample $z$, $t$ and proportion tests. Finally, Section S4 presents an instructional user guide for the R package. Designing and implementing a MSPRT, calculating the UMPBT alternative for different tests and obtaining $N^*$ (as in Section \ref{sec:Offsetting the cost of decreasing alpha}) are reviewed in respective subsections there. Additional simulation results for one-sample $z$ and proportion tests, 
	with similar conclusions as to the one-sample $t$ test, are presented in Section S4.2.
\end{sloppypar}

\section{Acknowledgment}
Support for this research was provided by the National Cancer Institute (R01 CA158113).


\bibliographystyle{apalike}

\bibliography{ref}


\end{document}


	
	
	
	\maketitle 
	
	
	\begin{abstract}
		Here we provide technical details for the main article. 
		We also provide a brief user's guidance to \texttt{MSPRT}, the \texttt{R} package that can be used to implement a MSPRT.
	\end{abstract}


\section{Introduction}

We consider null hypothesis significance tests (NHSTs) where the maximum number of samples ($N$) is specified and in which we wish to control Type I and Type II error probabilities at specified levels $\alpha $ and $\beta$, respectively. 

Let $X$ be a random variable having density $f({x} ; \theta)$ under both the null and alternative hypotheses, and let   $\theta$,  $\theta \in \Theta$, denote the parameter of interest.  Let $f({\bf x}_n;\theta)$ denote the joint sampling density of the observation ${\bf x}_n =\{x_1,\dots,x_n\}$ for some sample size $ n $, and let $\pi_i (\theta)$ denote the prior density assigned to $\theta$ under $H_i$ (for $i=0,1$). Then the marginal density $m_i ({\bf x}_n)$ of the data under $H_i$ (for $i=0,1$) is defined as
\begin{equation}\label{eq1}
m_i ({\bf x}_n) = \int_\Theta f({\bf x}_n ;\theta ) {\pi}_i (\theta) d\theta.
\end{equation}
For a given point alternative hypothesis $H_1:\theta = \theta_1$, we define the likelihood ratio (LR) as
\begin{equation}\label{eq2}
\LL = \frac{f({\bf x}_n; \theta_1 )}{f({\bf x}_n; \theta_0 )}.
\end{equation}
When there is no ambiguity regarding the values of $(\theta_0,\theta_1)$, we simply write $L_n \equiv \LL$. The Bayes factor (BF) in favor of $H_1$ is defined as ${BF}_{10} ({\bf x}_n) = \displaystyle{m_1 ({\bf x}_n)/m_0 ({\bf x}_n)}$. 

Following \cite{johnson13_2}, the uniformly most powerful Bayesian test (UMPBT) for evidence threshold $\delta >0$ in favor of the alternative $H_1$ against a fixed null $H_0$, denoted by UMPBT($\delta$), is a Bayesian hypothesis test in which the Bayes factor for the test satisfies the following inequality for any $\theta_t \in \Theta$ and for all alternative hypotheses $H_2 : \theta \sim \pi_2 (\theta)$:
\begin{equation}\label{eq3}
{\bf P}_{{\theta}_t} [{BF}_{10} ({\bf x}) > \delta] \geq {\bf P}_{{\theta}_t} [{BF}_{20} ({\bf x}) > \delta].
\end{equation}
That is, the UMPBT maximizes the probability that the Bayes factor against a fixed null hypothesis exceeds a specified threshold. Following equation (\ref{eq3}) for one-parameter exponential family models, the UMPBT alternative is defined as the alternative $\theta_1$ which maximizes ${\bf P}_{{\theta}_t}[{BF}_{10}({\bf x})>\delta]$ among all prior densities on $\theta$,  $\theta \in \Theta$. A list of the UMPBT alternatives for common statistical tests can be found in the supporting information file of \cite{johnson13_1}.

In tests of a simple null against a composite alternative, there is often a correspondence between the rejection regions of Bayesian testing rules using a UMPBT alternative and classical uniformly most powerful (UMP) tests (when such tests exist). Given a $ \delta $, the UMPBT($\delta$) alternative is optimal in the sense that it maximizes the probability that the Bayes factor in favor of the alternative exceeds a specified threshold $\delta$.  In such cases, $\delta$ can be determined by matching the rejection region of the test to that of the classical Neyman-Pearson UMP test of size $\alpha$. This naturally induces a one-to-one correspondence between the size of the test ($\alpha$) and the Bayesian evidence threshold ($\delta$).

In the rest of the discussion, we refer to the UMP test as the fixed-design test.

\medskip
\section{The Modified Sequential Probability Ratio Test (MSPRT)}\hypertarget{S2}{}

Given $N$, $\alpha$, and $\beta$, suppose we are interested in testing a simple null against a one-sided alternative, i.e.,  
\begin{equation}\label{eq4}
H_0 : \theta = \theta_0 \quad \text{vs.} \quad H_1 : \theta > \theta_0 \quad \text{or} \quad \theta < \theta_0,
\end{equation}
where $\theta$ is a scalar parameter defining $ f(x;\theta) $. We further assume that $ f(x;\theta) $ belongs to a one-parameter exponential family. Then, following the preceding discussion, we can obtain the UMPBT alternative by matching the UMPBT's rejection region to that of the fixed-design test using $N$ samples. Doing so leads to the definition of the UMPBT alternative hypothesis and the evidence threshold. Once the alternative is determined, we can compute the likelihood ratio (or Bayes factor) in favor of the alternative as we observe data sequentially.  For each $ n $, let $ L_n $ denote the likelihood ratio as defined in equation (1) in the main article. As in the case of SPRTs, we define the acceptance and rejection threshold for $ L_n $ by $B = \dfrac{\beta}{1 - \alpha}$ and $A = \dfrac{1- \beta}{\alpha}$, respectively. Using this notation, the conduct of the MSPRT can be defined by Algorithm \ref{algo1}.

\medskip
\begin{algorithm}
	\caption{\textbf{: MSPRT}}\label{algo1}
	\vspace{2mm}
	For $n=1, \ldots, N$
	
	\quad 1. \textbf{Stop} and \textbf{reject} $H_0$ \textbf{if} $L_n \geq A$
	
	\quad 2. \textbf{Stop} and \textbf{accept} $H_0$ \textbf{if} $L_n \leq B$
	
	\quad 3. \textbf{Collect the next} data point \textbf{if} $ B < L_n < A$
	
	\vspace{2mm}
	\noindent If \textbf{no decision} has been made after collecting $\bs{N}$ observations, \textbf{terminate} the procedure and \textbf{reject} $H_0$ \textbf{if} $L_N \geq \gamma$; otherwise, \textbf{accept} $H_0$.
\end{algorithm}

The threshold $\gamma$, which we refer to as the termination threshold, is chosen to be the smallest number that preserves the targeted size of the test $ \alpha $. In general, numerical procedures are required to determine the value of $\gamma$. We can implement this procedure using the \texttt{R} package \texttt{MSPRT}. A more detailed illustration for common tests is provided in Section~\hyperlink{S4}{S4}.

\section{Examples}

\medskip
\subsection{One-sample $z$ test for a population mean}\hypertarget{S3.1}{}

Suppose $X_1,\dots,X_N$ are  \textit{i.i.d.}~$N(\mu,\sigma^2)$ random variables, $\sigma^2$ is known, and we wish to test
\begin{equation}\label{eq5}
H_0 : \mu= {\mu}_0 \quad \text{vs.} \quad H_1 : \mu > {\mu}_0.
\end{equation}
Following \cite{johnson13_2}, the UMPBT($\delta$) alternative is defined as
\begin{equation}\label{eq6}
{\mu}_{1N} = \argminA_{\mu > {\mu}_0} \hspace{2mm} \displaystyle{ \Bigg [ \frac{ {\sigma}^2 \log \delta}{N(\mu - {\mu}_0)} + \frac{(\mu + {\mu}_0)}{2} \Bigg ] } = {\mu}_0 + \sigma \displaystyle{\sqrt{\frac{2 \log \delta}{N}}}.
\end{equation}

By matching the rejection region from the UMPBT with that of the fixed-design test, we obtain the evidence threshold as
\begin{equation}\label{eq7}
\delta = \displaystyle{\exp \Big( \frac{z^2_\alpha}{2} \Big)},
\end{equation}
where $z_{\alpha}$ is the ${100(1- \alpha)}$th quantile of the standard normal distribution. Substituting this in (\ref{eq6}), we get the UMPBT alternative
\begin{equation}\label{eq8}
\mu_{1N} = {\mu}_0 + z_{\alpha} \frac{\sigma}{\sqrt{N}}.
\end{equation}
The alternative corresponds to the rejection boundary for the fixed-design test of size $\alpha$ based on $N$ observations.

Using the alternative, we compute $L_n$ as
\begin{equation}\label{eq9}
L_n = \displaystyle{ \frac{f({\bf x}_{n};{\mu}_{1N})}{f({\bf x}_{n};\mu_0)}}= \displaystyle{ \exp \Bigg[ \frac{ ( {\mu}_{1N} - {\mu}_0 ) }{\sigma^2} {\sum\limits_{i=1}^n x_i} - \frac{n ( \mu^2_{1N} - \mu_0^2 ) }{2 \sigma^2} \Bigg] }.
\end{equation}

After $ \gamma $ is obtained, the MSPRT is then conducted according to Algorithm \ref{algo1} in Section~\hyperlink{S2}{S2}.\\

\medskip
\subsection{One-sample $t$ test for a population mean}\label{ttest}

Now suppose the conditions of Section~\hyperlink{S3.1}{S3.1} apply, but $\sigma^2$ is not known. 

A UMPBT does not exist in this case.  For this reason, we instead use the approximate data-dependent UMPBT($\delta$) alternative defined in \cite{johnson13_1} as 
\begin{equation}\label{eq10}
{\mu}_{1N} = {\mu}_0 + s_N \displaystyle{\sqrt{\frac{ \nu {\delta}^*}{N}}}
\end{equation}
\noindent where $s^2_N = \displaystyle{\frac{1}{N-1} \sum\limits_{i=1}^N (x_i - \bar{x}_N)^2 }$, $\nu = N-1$, and $\displaystyle{{\delta}^* = {\delta}^{2/N} -1}$.

Based on the maximum sample size $N$, the condition for matching the rejection regions of the UMPBT and the fixed-design $t$ test can be derived as
\begin{equation}\label{eq11}
\sqrt{ \nu {\delta}^*} = t^2_{\alpha;N-1}  \qquad  \Leftrightarrow \qquad \delta =  {\Bigg[ \frac{t^2_{\alpha;N-1}}{\nu} +1 \Bigg]}^{\displaystyle{\frac{N}{2}}},
\end{equation}
where $t_{\alpha;N-1}$ is the ${100(1- \alpha)}$th quantile of a $t$ distribution with degrees of freedom (df) $N-1$.

From observed data, we obtain the UMPBT alternative at step $n$ as
\begin{equation}\label{eq12}
\mu_{1n} = {\mu}_0 + t_{\alpha;N-1} \frac{s_n}{\sqrt{N}},
\end{equation}
for $n=2,...,N$.

Using this alternative, we define the integrated likelihood function (or Bayes factor) $L_n$ according to
\begin{equation}\label{eq13}
L_n = \displaystyle{ {\Bigg[ \frac{1+ \big( \frac{n}{n-1} \big) t^2_{0,n} }{1+ \big( \frac{n}{n-1} \big) t^2_{1,n} } \Bigg]}^{\displaystyle{ \frac{n}{2}} } },
\end{equation}
where $ t_{0,n} = \dfrac{ {\bar{x}}_n - \mu_0}{s_n} $ and $ t_{1,n} = \dfrac{ {\bar{x}}_n - \mu_{1n} }{s_n} $.

We obtained this integrated likelihood by imposing the noninformative prior $\pi(\sigma^2) \propto 1/\sigma^2$ on the unknown variance parameter. 

Once $ \gamma $ is determined numerically, the MSPRT is conducted according to Algorithm \ref{algo1} in Section~\hyperlink{S2}{S2}.\\

\medskip
\subsection{One-sample test for a binomial proportion}\label{proptest}

Suppose $X_1,\dots,X_N$ represent \textit{i.i.d.} Bernoulli observations with success probability $p$, and for some $ p_0 $ we wish to test 
\begin{equation}\label{eq14}
H_0 : p= p_0 \quad \text{vs.} \quad H_1 : p > p_0.
\end{equation}

To design the MSPRT, we must determine the alternative hypothesis that will be used to compute $L_n$. We can accomplish this most easily by first examining the form of the fixed design test's rejection region.  Based on the maximum sample size $N$, that test rejects $H_0$ if
\begin{equation}\label{eq15}
\sum_{i=1}^N X_i > c_0,
\end{equation}
where $c_0$ is defined as
\begin{equation}\label{eq16}
c_0 = \inf \bigg \{ c \bigg| P_{H_0}\bigg( \sum\limits_{i=1}^N x_i > c \bigg) \leq \alpha \bigg \}.
\end{equation}

Following \cite{johnson13_2}, the UMPBT($\delta$) alternative value of $p$ is defined as	
\begin{equation}\label{eq17}
	p_{1N} (\delta) =  \argminA_{p> p_0} \hspace{2mm} h_N (p, \delta),
\end{equation}
where
\begin{equation}\label{eq18}
	h_N (p, \delta) =  \frac{ \log \delta - N \bigg[ \log(1-p)- \log(1- p_0) \bigg]}{\displaystyle{ \log\bigg(\frac{p}{1-p}\bigg) - \log\bigg(\frac{p_0}{1- p_0}}\bigg)} .
\end{equation}
For a given $(p,\delta)$, the rejection region for the UMPBT($\delta$) test is
\begin{equation}\label{eq19}
\sum_{i=1}^N X_i > h_N(p, \delta).
\end{equation}
Thus, the rejection region from the fixed-design test can be matched to that of the UMPBT by solving
\begin{equation}\label{eq20}
h_N \Big( p_{1N}(\delta), \delta \Big)= c_0
\end{equation} 
for $ \delta $. This solution provides the evidence threshold for the test.
	



In practice, the discrete nature of binomial data causes  the Type I error of the test to be less than the targeted $\alpha$.  In order to achieve the exact $ \alpha $ in a classical test, one must use a randomized test. The randomized test can be described as follows: with probability $\psi$, reject $ H_0 $ if $ \sum\limits_{i=1}^N x_i > (c_0 -1) $, and with probability $ (1-\psi )$, reject $H_0$  if $ \sum\limits_{i=1}^N x_i > c_0 $.  The value of $\psi$ is determined according to
\begin{equation}\label{eq21}
\psi = \bigg[ \alpha - P_{H_0}\bigg( \sum\limits_{i=1}^N x_i > c_0 \bigg) \bigg] \bigg/ P_{H_0}\bigg( \sum\limits_{i=1}^N x_i = c_0 \bigg).
\end{equation}

This suggests that we obtain the UMPBT alternative according to the following modification. Noting that the fixed-design randomized test involves two rejection regions, namely $ \Big( c_0 -1 , N \Big] $ and $ \Big( c_0 , N \Big] $, and recalling (\ref{eq6}), we solve
\begin{equation}\label{eq22}
h_N \Big( p_{1N}(\delta_L), \delta_L \Big) = c_0 -1 \quad \text{and} \quad h_N \Big( p_{1N}(\delta_U), \delta_U \Big) = c_0.
\end{equation}
In contrast to {\em z} and {\em t} tests, using these values we define the UMPBT alternative as a mixture distribution of two points $ p_{1N}(\delta_L) \equiv p_{1N,L} $ and $ p_{1N}(\delta_U) \equiv p_{1N,U} $ with mixing probabilities $ \psi $ and $ (1- \psi) $, respectively. Then we obtain $ L_n $ as a weighted likelihood function defined by
\begin{equation}\label{eq23}
L_n = \psi \displaystyle{\frac{f( {\bf x}_{n} ; p_{1N,L})}{f( {\bf x}_{n} ; p_0)}} + (1- \psi) \frac{f( {\bf x}_{n} ; p_{1N,U})}{f( {\bf x}_{n} ; p_0)},
\end{equation}
where
\begin{equation}\label{eq24}
\frac{f( {\bf x}_{n} ; p )}{f( {\bf x}_{n} ; p_0)}= {\bigg[ \frac{1- p }{1-p_0} \bigg]}^n {\bigg[ \frac{p (1- p_0)}{p_0 (1-p)} \bigg]}^{\sum\limits_{i=1}^n x_i}.
\end{equation}

After $ \gamma $ has been numerically obtained, the MSPRT can be implemented using Algorithm \ref{algo1} in Section~\hyperlink{S2}{S2}.

\medskip
\section{Examples with \texttt{MSPRT}: A user's guide}\hypertarget{S4}{}

We have created an \texttt{R} package named \texttt{MSPRT} to implement the MSPRT conveniently.  We illustrate the use of the test in the following examples.  We assume throughout that \texttt{MSPRT} has been loaded into the \texttt{R} command environment.

\medskip
\subsection{Designing and implementing a MSPRT}

A key function in the package is \texttt{design.MSPRT()}. Given $N$, $\alpha$, $\beta$, and other parameters, this function finds the MSPRT. Recall from Algorithm \ref{algo1} that finding the MSPRT requires finding the termination threshold $ \gamma $. The function \texttt{design.MSPRT()} does this. It also provides an option (through the argument \texttt{theta1}) to find the performance of the resulting MSPRT at a user-defined point alternative.\\

\medskip
\subsubsection{One-sample $z$ test for a population mean}\hypertarget{S4.1.1}{}

Our first illustration of the MSPRT is for a simple {\em z} test.  For concreteness, suppose we wish to test $ H_0: \mu=3 $ against the alternative hypothesis $H_1: \mu>3$ for a fixed $ \sigma=1.5 $ with a maximum of $ N=30 $ patients in a $ \alpha = 0.5\% $ test with Type II error of approximately $ \beta=0.2 $. There are two steps in the testing process: design and implementation.

In the design step, we calculate the termination threshold and the operating characteristics of the MSPRT. To do this, we use the functions \texttt{design.MSPRT()} and \texttt{OCandASN.MSPRT()}, respectively. The function \texttt{design.MSPRT()} is used to determine the termination threshold and evaluate the performance of the MSPRT when the null hypothesis is true. The required commands follow:

\medskip
{\small
\begin{verbatim}
> design.out = design.MSPRT(test.type = "oneZ", theta0 = 3, sigma = 1.5, N.max = 30)
> design.out$TypeI.attained    ## Type I error probability
[1] 0.005
> design.out$EN[1]    ## avg. sample size under the null
[1] 14.24063
> design.out$theta.UMPBT    ## UMPBT alternative
[1] 3.70542
> design.out$termination.threshold    ## termination threshold
[1] 27.911
\end{verbatim}
}

In this code snippet, the values \texttt{TypeI.attained}, \texttt{EN[1]}, and \texttt{termination.threshold} represent the Type I error probability, the average sample size required for reaching a decision when the null hypothesis is true, and the termination threshold of the  MSPRT, respectively.

Normally, we must find the operating characteristics of the test at several alternative values.  For the UMPBT alternative (equal to 3.7054 in this case), these values can be obtained by giving the following command.

\medskip
{\small
	\begin{verbatim}
	> OC.out = OCandASN.MSPRT(theta = 3.7054, design.MSPRT.object = design.out)
	> OC.out$acceptH0.prob   ##Type II error at the UMPBT alternative
[1] 0.509086
> OC.out$EN   ##avg. sample size at the UMPBT alternative
[1] 25.29154
	\end{verbatim}
}

The values returned from this function call include (but are not restricted to) \texttt{acceptH0.prob} and \texttt{EN}. They are interpreted as the Type II error probability and the average sample size required by the designed 
MSPRT for reaching a decision when the UMPBT alternative is true, respectively.

Finally, it may be necessary to obtain the operating characteristics at arbitrary values of the alternative hypothesis. Again for concreteness, suppose we wish to determine the operating characteristics for $ \mu=4 $ (for example). Then the following command may be given.  

\medskip
{\small
	\begin{verbatim}
	> OC.out = OCandASN.MSPRT(theta = 4, design.MSPRT.object = design.out)
> OC.out$acceptH0.prob   ##Type II error at the the desired alternative
[1] 0.151229
> OC.out$EN   ##avg. sample size at the desired alternative
[1] 22.67337
	\end{verbatim}
}

The output from this command may be interpreted as before.

Next, in the implementation phase we can apply the test to a sequence of observed values.  To illustrate this procedure, we simulate the observed values as follows:

\medskip
{\small
	\begin{verbatim}
	> set.seed(1)
> x = rnorm(n = 30, mean = 5, sd = 1.5)
	\end{verbatim}
}

Given these values, the MSPRT stopping criteria can be tested with the command \texttt{implement.MSPRT()}. Note that the object \texttt{design.out} is obtained using the \texttt{design.MSPRT()} command as above.

\medskip
{\small
	\begin{verbatim}
	> implement.out = implement.MSPRT(obs = x, design.MSPRT.object = design.out)
	> implement.out$decision   ##decision
	[1]  "reject.null"
	> implement.out$n   ##number of observations required to reach the decision
	[1]  9
	\end{verbatim}
}

This output shows that the null hypothesis is rejected after the $9^{th}$ observation.

If \texttt{plot.it = 2} (the default), the call to \texttt{implement.MSPRT()} also returns a sequential comparison plot similar to that depicted in Figure~\hyperlink{figS1}{S1}.  This particular plot shows that $L_n$ crosses the ``reject null" threshold on the $9^{th}$ observation, at which time the null hypothesis is rejected.

\begin{figure}[h]\hypertarget{figS1}{}
	\centering
	\includegraphics[width=0.8\textwidth]{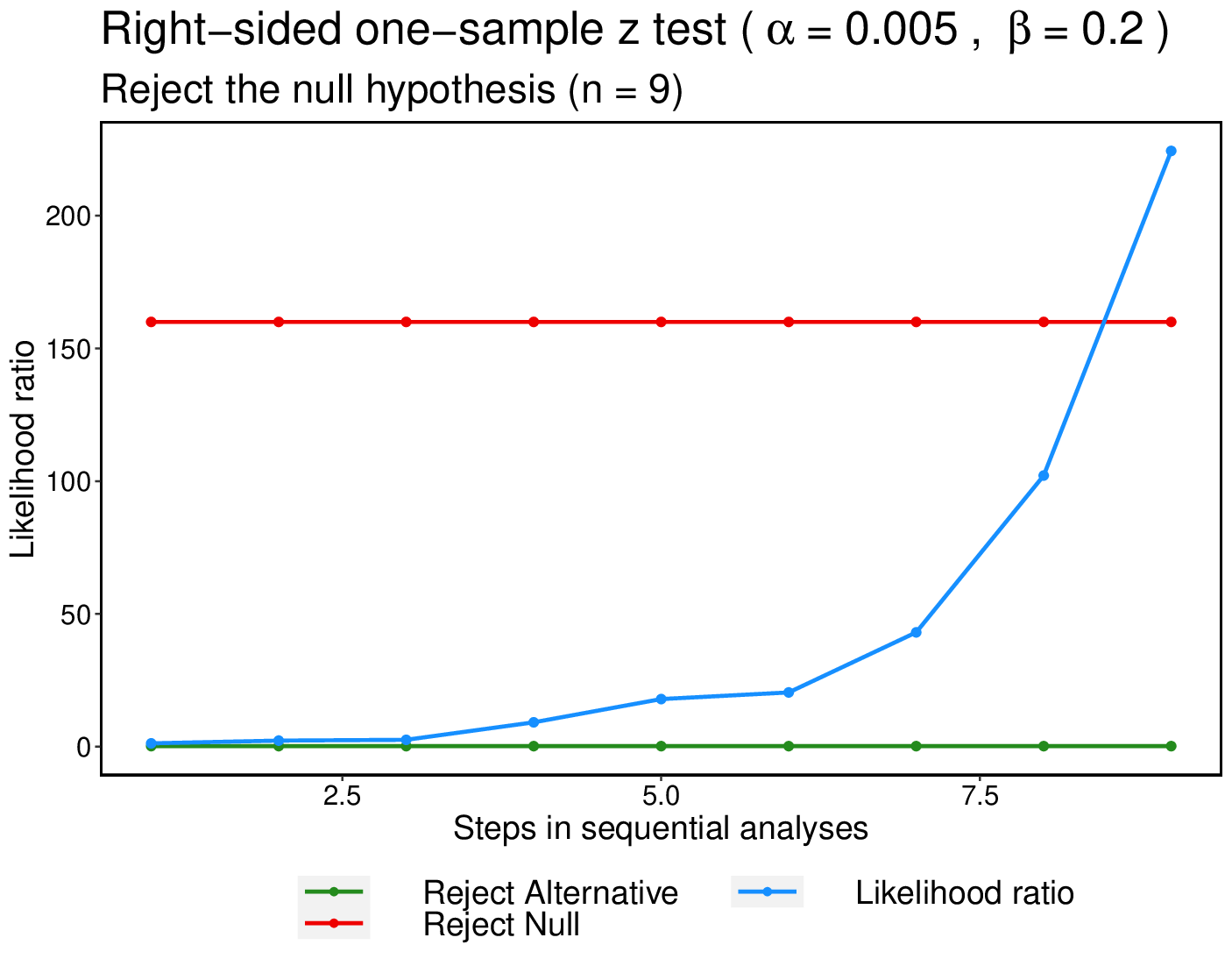}
	\caption*{{{\em Figure S1.} { One-sample} {\em z} test that a population mean equals 3.}  Hypothesis test of $ H_0 : \mu = 3 $ vs. $ H_1 : \mu > 3 $ with $ \sigma $ known to be 1.5. Sequential comparison plot of the MSPRT obtained in Section~\hyperlink{S4.1.1}{S4.1.1}.}
\end{figure}

\medskip
\subsubsection{One-sample $t$ test for a population mean}\hypertarget{S4.1.2}{}

Our next illustration of the MSPRT is for a {\em t} test.  For concreteness, suppose we again wish to test $ H_0: \mu=3 $ against an alternative hypothesis $H_1: \mu >3$ for an unknown $ \sigma $ with a maximum of $ N=30 $ patients in a $ \alpha = 0.5\% $ test with Type II error of approximately $ \beta=0.2 $. Again there are two steps in the testing process: design and implementation.

In the design step, we calculate the termination threshold and the operating characteristics of the MSPRT.  To do this, we again use the functions \texttt{design.MSPRT()} and \texttt{OCandASN.MSPRT()}, respectively. The function \texttt{design.MSPRT()} is used to determine the termination threshold and evaluate the performance of the MSPRT when the null hypothesis is true. The required commands follow:

\medskip
{\small
	\begin{verbatim}
	> design.out = design.MSPRT(test.type = "oneT", theta0 = 3, N.max = 30)
> design.out$TypeI.attained    ## Type I error probability
[1] 0.005
> design.out$EN[1]    ## avg. sample size under the null
[1] 14.60748
> design.out$termination.threshold    ## termination threshold
[1] 34.02
	\end{verbatim}
}

The values \texttt{TypeI.attained}, \texttt{EN[1]}, and 
\texttt{termination.threshold} can be interpreted as before.

Once we have obtained the MSPRT design, it may be necessary to obtain the operating characteristics of the test at arbitrary values of the alternative hypothesis. Again for concreteness, suppose we wish to determine the operating characteristics for $ \mu = 4 $. We can do that by using the following command.

\medskip
{\small
	\begin{verbatim}
	> OC.out = OCandASN.MSPRT(theta = 4, design.MSPRT.object = design.out)
> OC.out$acceptH0.prob   ##Type II error at the the desired alternative
[1] 0.006113
> OC.out$EN   ##avg. sample size at the desired alternative
[1] 22.39615
	\end{verbatim}
}

The values can be interpreted as in the previous section.

Next, in the implementation phase we can apply the test to a sequence of observed values.  To illustrate this procedure, we use the same \texttt{x} as in Section~\hyperlink{S4.1.1}{S4.1.1}:

\medskip
{\small
	\begin{verbatim}
	> set.seed(1)
> x = rnorm(n = 30, mean = 5, sd = 1.5)
	\end{verbatim}
}

Given these values, the MSPRT stopping criteria can be tested with the command \texttt{implement.MSPRT()}. Note that the object \texttt{design.out} is obtained using the \texttt{design.MSPRT()} command as above.

\medskip
{\small
	\begin{verbatim}
	> implement.out = implement.MSPRT(obs = x, design.MSPRT.object = design.out)
	> implement.out$decision   ##decision
	[1]  "reject.null"
	> implement.out$n   ##number of observations required to reach decision
	[1]  22
	\end{verbatim}
}

Output from these commands shows that the null hypothesis is rejected after the ${22}^{nd}$ observation.

If \texttt{plot.it = 2} (the default), the call to \texttt{implement.MSPRT()} also returns a sequential comparison plot similar to that depicted in Figure~\hyperlink{figS2}{S2}. This particular plot show that $L_n$ crosses the ``reject null" threshold on the ${22}^{nd}$ observation, at which time the null hypothesis is rejected.

\begin{figure}[h]\hypertarget{figS2}{}
	\centering
	\includegraphics[width=0.8\textwidth]{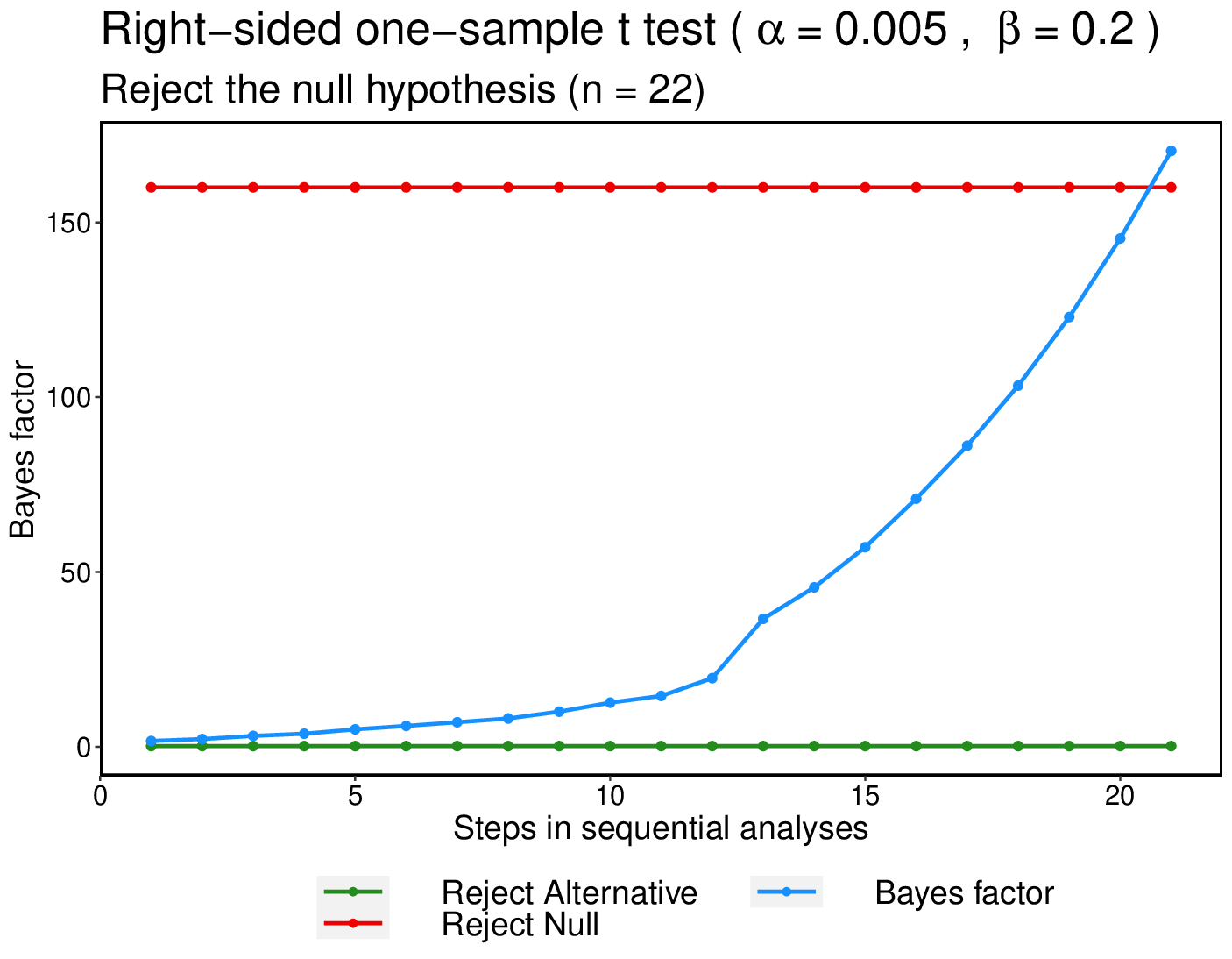}
	\caption*{{{\em Figure S2.} { One-sample} {\em t} test that a population mean equals 3.}  Hypothesis test of $ H_0 : \mu = 3 $ vs. $ H_1 : \mu > 3 $ when $ \sigma $ is assumed unknown. Sequential comparison plot of the MSPRT obtained in Section~\hyperlink{S4.1.2}{S4.1.2}.}
	\label{ttestfig}
\end{figure}

\medskip
\subsubsection{One-sample test of a binomial proportion}\hypertarget{S4.1.3}{}

We next consider the MSPRT for a proportion test. For concreteness, suppose we wish to test $H_0: p=0.2 $ against the alternative hypothesis $H_1: p>0.2$  with a maximum of $ N=30 $ patients in a $ \alpha = 0.5\% $ test with Type II error of approximately $ \beta=0.2 $. Again we go through the two steps in the testing process: design and implementation.

In the design step, we calculate the termination threshold and the operating characteristics of the MSPRT.  To do this, we again use functions \texttt{design.MSPRT()} and \texttt{OCandASN.MSPRT()}, respectively.  The commands follow:

\medskip
{\small
	\begin{verbatim}
	> design.out = design.MSPRT(test.type = "oneProp", theta0 = 0.2, N.max = 30)
	> design.out$Type1.attained   ##Type I error probability
[1] 0.002946
> design.out$EN[1]   ##avg. sample size under the null
[1] 12.9514
> design.out$UMPBT$theta   ##two points of the UMPBT alternative
[1] 0.3666727 0.4000178
> design.out$UMPBT$mix.prob   ##mixing probability for the UMPBT alternative
[1] 0.2959777 0.7040223
> design.out$termination.threshold   ##termination threshold
[1] 13.21
	\end{verbatim}
}

The values \texttt{TypeI.attained}, \texttt{EN[1]}, and \texttt{termination.threshold} can be interpreted as before. The values of \texttt{UMPBT\$theta} and \texttt{UMPBT\$mix.prob} together specify the UMPBT alternative used by the MSPRT. In this case the alternative is 0.3667 and 0.4 with approximate probabilities 0.296 and 0.704, respectively.

Once we have the MSPRT design, we can use \texttt{OCandASN.MSPRT()} to compute the operating characteristics of that MSPRT.  For concreteness, suppose we wish to determine the operating characteristics for $ p = 0.3 $. The following commands do this.

\medskip
{\small
	\begin{verbatim}
	> OC.out = OCandASN.MSPRT(theta = 0.3, design.MSPRT.object = design.out)
> OC.out$acceptH0.prob   ##Type II error at the the desired alternative
[1] 0.920718
> OC.out$EN   ##avg. sample size at the desired alternative
[1] 20.1515
	\end{verbatim}
}

The values returned from this function call have the same interpretation as before.

Next, in the implementation phase we can apply the test to a sequence of observed values. To illustrate this procedure, we simulate the observed binary values as follows:

\medskip
{\small
	\begin{verbatim}
	> set.seed(1)
> x = rbinom(n = 30, size = 1, prob = 0.2)
	\end{verbatim}
}

Given these values, the MSPRT stopping criteria can be tested with the command \texttt{implement.MSPRT()}. Note that the object \texttt{design.out} is obtained using the \texttt{design.MSPRT()} command as above.

\medskip
{\small
	\begin{verbatim}
	> implement.out = implement.MSPRT(obs = x, design.MSPRT.object = design.out)
	> implement.out$decision   ##decision
[1] "reject.alt"
> implement.out$n   ##number of observations required to reach decision
[1] 15
	\end{verbatim}
}

This output shows that the alternative hypothesis is rejected after using the $15^{th}$ observation.
In particular, the sequential test plot in Figure~\hyperlink{figS3}{S3} shows the sequential trajectory of $L_n$ until the alternative hypothesis is rejected.

\begin{figure}[h]\hypertarget{figS3}{}
	\centering
	\includegraphics[width=0.8\textwidth]{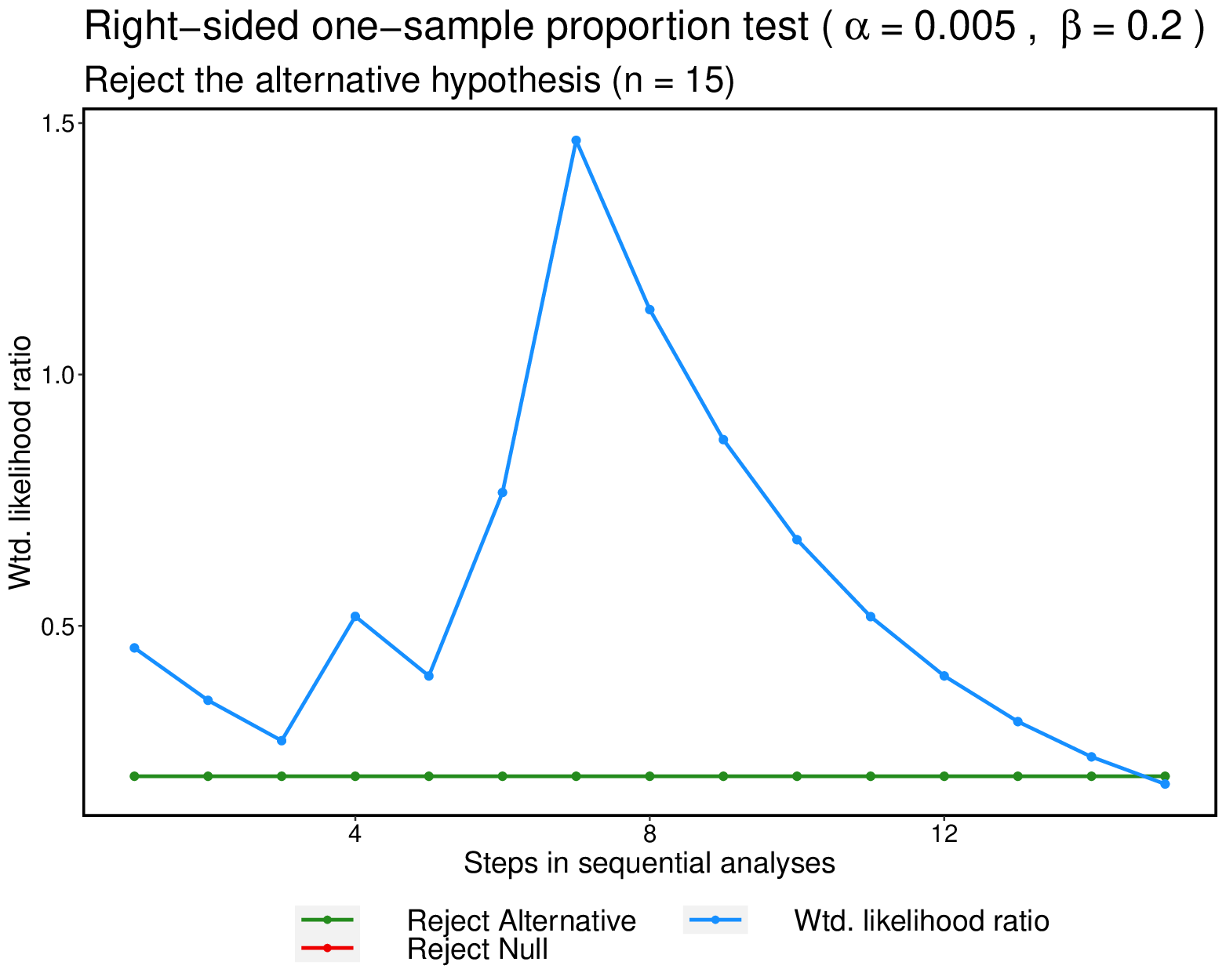}
	\caption*{{{\em Figure S3.} { One-sample} test that a binomial proportion equals 0.2.}  Hypothesis test of $ H_0 : p = 0.2 $ vs. $ H_1 : p > 0.2 $ . Sequential comparison plot of the MSPRT as in Section~\hyperlink{S4.1.3}{S4.1.3}.}
	\label{proptestfig}
\end{figure}

{
\medskip
\subsubsection{Two-sample $z$ test for a difference in two population means}\hypertarget{S4.1.4}{}

Let, $ \mu_1 $ and $ \mu_2 $ be the population means of two groups of patients, respectively. Suppose we want to test $ H_0: \mu_1 - \mu_2 =0 $ against the alternative hypothesis $H_1: \mu_1 - \mu_2 >0$ for a known common population variance of $ \sigma = 1.5 $.  Assume that we can observe a maximum of 30 patients from each group (that is, $ N_1 = N_2 = 30 $). We set $ \alpha = 0.5\% $ and the Type II error level $ \beta=0.2 $. 

In the design step, we calculate the termination threshold and the operating characteristics of the MSPRT.  As before, the function \texttt{design.MSPRT()} is used to determine the termination threshold and evaluate the performance of the MSPRT when the null hypothesis is true. The required commands are as follows:

\medskip
{\small
	\begin{verbatim}
	> design.out = design.MSPRT(test.type = "twoZ", sigma1 = 1.5, sigma2 = 1.5, 
	                            N1.max = 30, N2.max = 30)
	> design.out$Type1.attained   ##Type 1 error probability
	[1] 0.005
	> design.out$EN$H0
	$Group1   ##avg. sample size from Group 1 under the null
	[1] 14.22938
	
	$Group2   ##avg. sample size from Group 2 under the null
	[1] 14.22938
	> design.out$theta.UMPBT   ##UMPBT alternative
	[1] 0.9976144
	> design.out$termination.threshold   ##termination threshold
	[1] 27.885
	\end{verbatim}
}

In this code snippet, the values \texttt{TypeI.attained}, \texttt{EN\$H0}, and \texttt{termination.threshold} respectively represent the Type I error probability, the average sample size required from Group 1 and 2 under the null hypothesis, and the termination threshold of the designed 
MSPRT.

Normally, we must also find the operating characteristics of the test at several alternative values.  For the UMPBT alternative (equal to 0.9976 in this case), these values can be obtained by giving the following command.

\medskip
{\small
	\begin{verbatim}
	> OC.out = OCandASN.MSPRT(theta = 0.9976144, design.MSPRT.object = design.out)
	> OC.OC.out$acceptH0.prob   ##Type II error at the UMPBT alternative
	[1] 0.509531
	> OC.out$EN1   ##avg. sample size from Group 1 at the UMPBT alternative
	[1] 25.31669
	> OC.out$EN2   ##avg. sample size from Group 2 at the UMPBT alternative
	[1] 25.31669
	\end{verbatim}
}

The values returned from this function call include \texttt{theta},  \texttt{acceptH0.prob}, \texttt{EN1}, and \texttt{EN1}. They are interpreted as the effect size where the performace is evaluated, the Type II error probability, the average sample size required from Group 1 at the UMPBT alternative, and the average sample size required from Group 2 at the UMPBT alternative, respectively.

To obtain the operating characteristics at arbitrary values of the alternative hypothesis, suppose we wish to determine the operating characteristics for $ \mu_1 - \mu_2 = 2 $. Then the following command may be given.  

\medskip
{\small
	\begin{verbatim}
	> OC.out = OCandASN.MSPRT(theta = 2, design.MSPRT.object = design.out)
	> OC.out$acceptH0.prob   ##Type II error at the desired alternative
	[1] 0.007961
	> OC.out$EN1   ##avg. sample size from Group 1 at the desired alternative
	[1] 16.17953
	> OC.out$EN2   ##avg. sample size from Group 2 at the desired alternative
	[1] 16.17953
	\end{verbatim}
}

The output from this command may be interpreted as before.

Next, in the implementation phase we can apply the test to two sequences of observed values from both groups.  To illustrate this procedure, suppose that we simulate the observed values from Group 1 and 2 as follows:

\medskip
{\small
	\begin{verbatim}
	> set.seed(1)
	> x1 = rnorm(n = 30, mean = 0.998, sd = 1.5)
	> x2 = rnorm(n = 30, mean = 0, sd = 1.5)
	\end{verbatim}
}

Given these values, the MSPRT stopping criteria can be tested with the command \texttt{implement.MSPRT()}. Note that the object \texttt{design.out} is obtained using the \texttt{design.MSPRT()} command as above.

\medskip
{\small
	\begin{verbatim}
	> implement.out = implement.MSPRT(obs1 = x1, obs2 = x2, 
	                                  design.MSPRT.object = design.out)
	> implement.out$decision   ##decision
	[1] "reject.alt"
	> implement.out$n1   ##number of observations required from Group 1
	[1] 30
	> implement.out$n2   ##number of observations required from Group 2
	[1] 30
	\end{verbatim}
}

This output shows that the alternative hypothesis is rejected after using the maximum number of available samples from each group.

If \texttt{plot.it = 2} (the default), the call to \texttt{implement.MSPRT()} also returns a sequential comparison plot similar to that depicted in Figure~\hyperlink{figS4}{S4}. This particular plot shows that $L_n$ reaches $N = 30$ without reaching a decision. But the likelihood ratio at the maximum sample size is approximately $L_{30} = 16.74$ (stored in \texttt{implement.out\$LR}), which is below the termination threshold 27.885. So the test rejects the alternative after observing 30 samples from each group.

\begin{figure}[h]\hypertarget{figS4}{}
	\centering
	\includegraphics[width=0.8\textwidth]{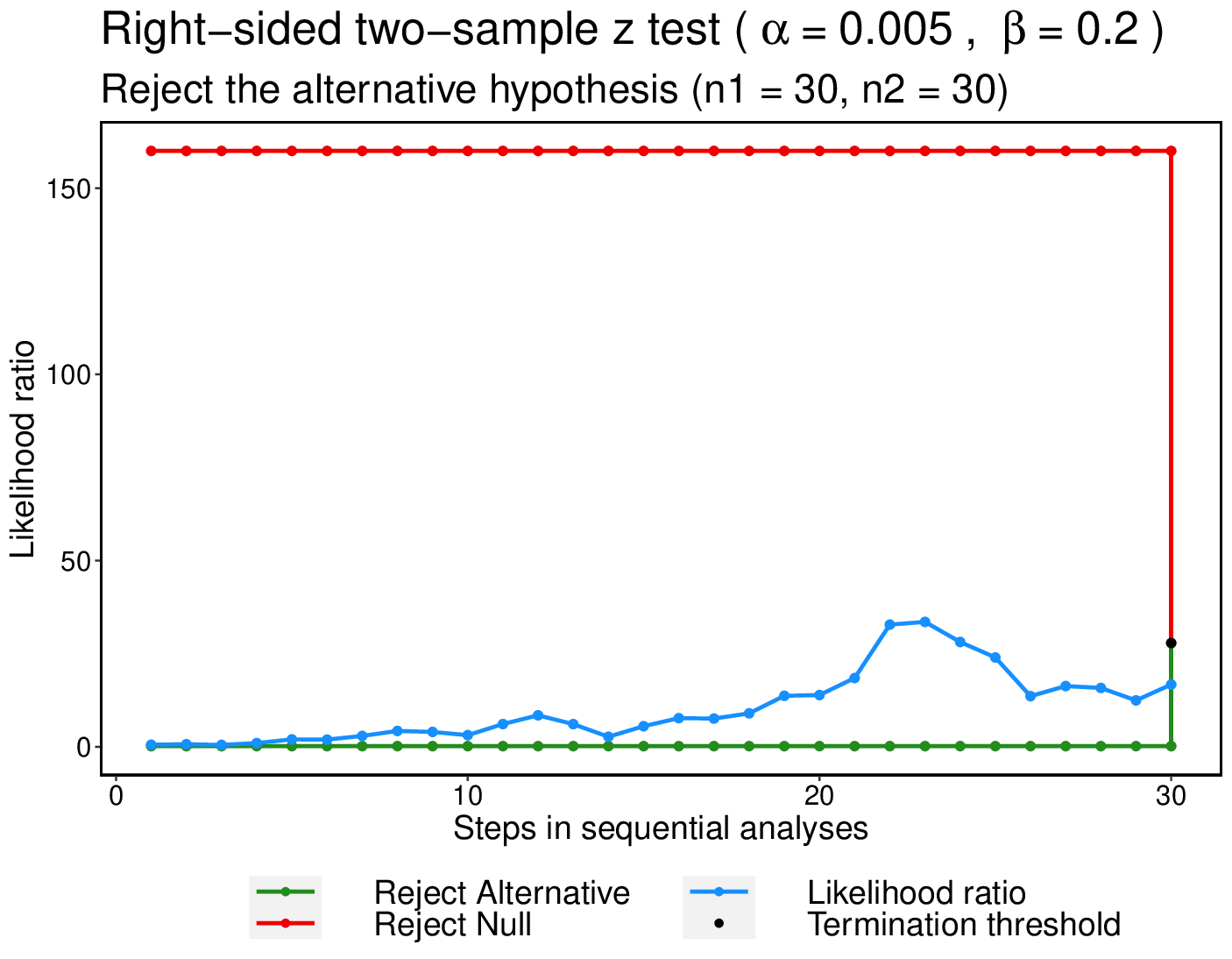}
	\caption*{{\em Figure S4.} Two-sample {\em z} test that the difference in population means is 0.  Hypothesis test of $ H_0 : \mu_1 - \mu_2 = 0 $ vs. $ H_1 : \mu_1 - \mu_2 > 0 $ with known common population standard deviation $1.5$. Sequential comparison plot of the MSPRT obtained in Section~\hyperlink{S4.1.4}{S4.1.4}.}
	\label{twoztestfig}
\end{figure}

\medskip
\subsubsection{Two-sample $t$ test for a difference in two population means}\hypertarget{S4.1.5}{}

Assume the exact setup as in \hyperlink{S4.1.4}{S4.1.4}, and suppose we want to test $ H_0: \mu_1 - \mu_2 =0 $ against $H_1: \mu_1 - \mu_2 >0$, but the common population variance is unknown. 

In the design step, we calculate the termination threshold and the operating characteristics of the MSPRT.  The required commands follow:

\medskip
{\small
	\begin{verbatim}
	> design.out = design.MSPRT(test.type = "twoT", N1.max = 30, N2.max = 30)
	> design.out$Type1.attained   ##Type 1 error probability
	[1] 0.005
	> design.out$EN$H0
	$Group1   ##avg. sample size from Group 1 under the null
	[1] 13.93484
	$Group2   ##avg. sample size from Group 2 under the null
	[1] 13.93484
	> design.out$termination.threshold   ##termination threshold
	[1] 33.243
	\end{verbatim}
}

In this code snippet, the values \texttt{Type1.attained}, \texttt{EN\$H0}, and \texttt{termination.threshold} represent the Type I error probability, the average sample size required from each group under the null hypothesis, and the termination threshold of the designed 
MSPRT, respectively.

To obtain the operating characteristics at arbitrary values of the alternative hypothesis, say, $ \mu_1 - \mu_2 = 2 $, the following command may be given.  

\medskip
{\small
	\begin{verbatim}
	> OC.out = OCandASN.MSPRT(theta = 2, design.MSPRT.object = design.out)
	> OC.out$acceptH0.prob   ##Type II error at the UMPBT alternative
	[1] 4.9e-05
	> OC.out$EN1   ##avg. sample size from Group 1 at the desired alternative
	[1] 15.61961
	> OC.out$EN2   ##avg. sample size from Group 2 at the desired alternative
	[1] 15.61961
	\end{verbatim}
}

The output from this command may be interpreted as before.

Next, in the implementation phase we can apply the test to two sequences of observed values from both groups.  To illustrate this procedure, we use the same \texttt{x1} and \texttt{x2} as in Section~\hyperlink{S4.1.4}{S4.1.4}:

\medskip
{\small
	\begin{verbatim}
	> set.seed(1)
	> x1 = rnorm(n = 30, mean = 0.998, sd = 1.5)
	> x2 = rnorm(n = 30, mean = 0, sd = 1.5)
	\end{verbatim}
}

Given these values, the MSPRT stopping criteria can be tested with the command \texttt{implement.MSPRT()}. Note that the value of \texttt{termination.threshold} is obtained using the \texttt{design.MSPRT()} command above.

\medskip
{\small
	\begin{verbatim}
	> implement.out = implement.MSPRT(obs1 = x1, obs2 = x2,
	                                  design.MSPRT.object = design.out)
	> implement.out$decision   ##decision
	[1] "reject.null"
	> implement.out$n1   ##number of observations required from Group 1
	[1] 30
	> implement.out$n2   ##number of observations required from Group 2
	[1] 30
	\end{verbatim}
}

This output shows that the null hypothesis is rejected after observing the maximum available number of 30 patients from each group.

If \texttt{plot.it = 2} (the default), the call to \texttt{implement.MSPRT()} also returns a sequential comparison plot similar to that depicted in Figure~\hyperlink{S5}{S5}.  This particular plot shows that $L_n$ reaches $N = 30$ without reaching a decision. But the likelihood ratio at the maximum sample size is approximately $L_{30} = 40.615$ (stored in \texttt{implement.out\$LR}), which is above the termination threshold 33.243. So the test rejects the null after observing 30 samples from each group.

\begin{figure}[h]\hypertarget{S5}{}
	\centering
	\includegraphics[width=0.8\textwidth]{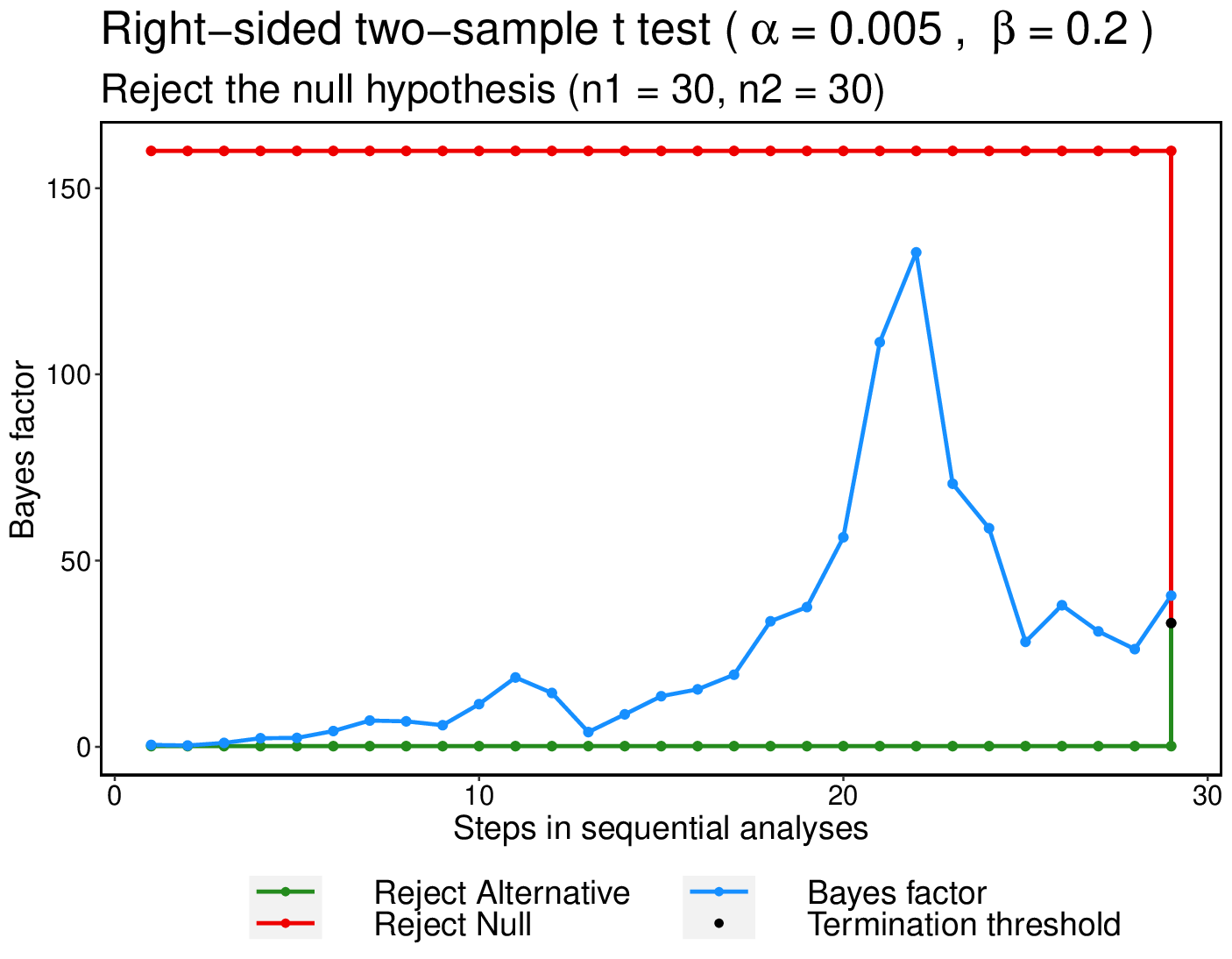}
	\caption*{{\em Figure S5.} Two-sample {\em t} test that the difference in population means is 0.  Hypothesis test of $ H_0 : \mu_1 - \mu_2 = 0 $ vs. $ H_1 : \mu_1 - \mu_2 > 0 $ with unknown common population standard deviation. Sequential comparison plot of the MSPRT obtained in Section~\hyperlink{S4.1.5}{S4.1.5}.}
	\label{twottestfig}
\end{figure}

}

{\color{black}
\medskip
\subsection{Results from simulation studies}

In this section we describe in more detail the simulation results from the main article. We examine one-sample tests for a binomial proportion, {\em z} tests and {\em t} tests of size $\alpha=0.05$ and $ 0.005 $. For simplicity, we examine one-sided tests with alternative hypotheses of the form $H_1: \theta>\theta_0$. We also assume that the targeted power of the test is 80\% (i.e., $\beta=0.2$).  Two-sided tests, tests of alternative hypotheses of the form $H_1: \theta<\theta_0$, and tests with different Type I or Type II errors are handled similarly. We compare the MSPRTs to standard fixed-design tests having the same size $\alpha$, sample size $N$, and Type II error $\beta=0.2$. {\color{black} Given $N$ and $\alpha$ for} 
fixed-design tests, we define $ \theta_a $, the fixed-design alternative, as the alternative parameter value {\color{black} (effect size)} that provides the specified $\beta$.

\begin{figure}[h]\hypertarget{S6}{}
	\centering
	\includegraphics[width=.8\textwidth]{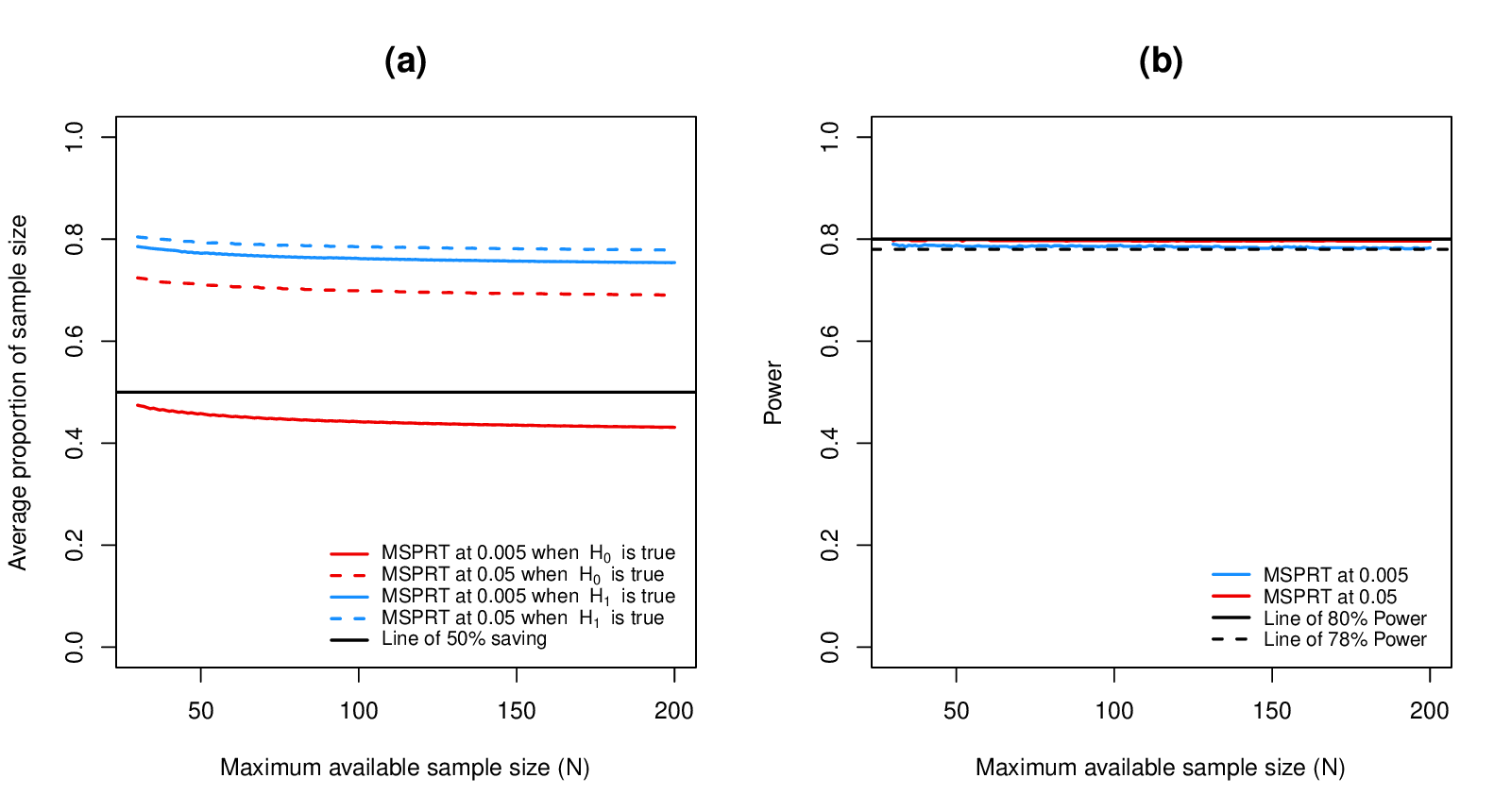}
	\caption*{{{\em Figure S6.} One-sample {\em z} test that a population mean equals 0.}  Hypothesis test of $ H_0 : \theta = 0$ vs. $ H_1 : \theta > 0 $.  The curves in the left plot represent the average proportion of the maximum sample size ($ N $) used before the MSPRT terminates in favor of the null or alternative hypothesis.  The plot on the right displays the average power of the test against its targeted value of 0.8. In both plots, the operating characteristics under the alternative are evaluated at the corresponding fixed-design alternatives. }\label{onez}
\end{figure}

\begin{figure}[h]\hypertarget{S7}{}
	\centering
	\includegraphics[width=.8\textwidth]{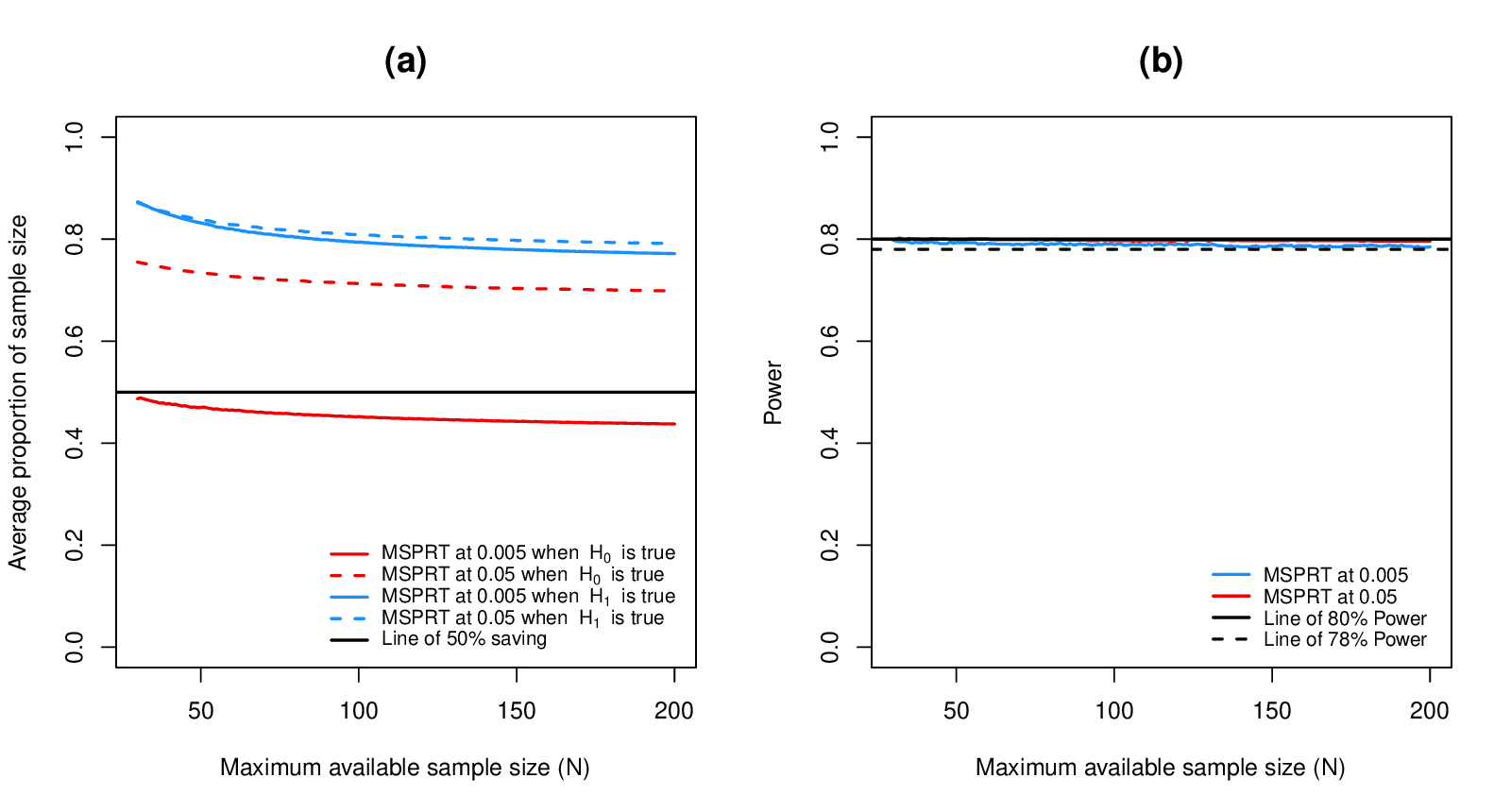}
	\caption*{{{\em Figure S7.} One-sample {\em t} test that a population mean is 0.} Hypothesis test of $ H_0 : \theta = 0$ vs. $ H_1 : \theta > 0 $. In contrast to Figure S6, the population standard deviation is assumed to be unknown. The curves in the left plot represent the average proportion of the maximum sample size ($N$) used before the MSPRT terminates in favor of the null or alternative hypothesis.  The plot on the right displays the average power of the test against its targeted value of 0.8.  In both plots, the operating characteristics under the alternative are evaluated at the corresponding fixed-design point alternatives. }\label{onet}
\end{figure}

\begin{figure}[h]\hypertarget{S8}{}
	\centering
	\includegraphics[width=.8\textwidth]{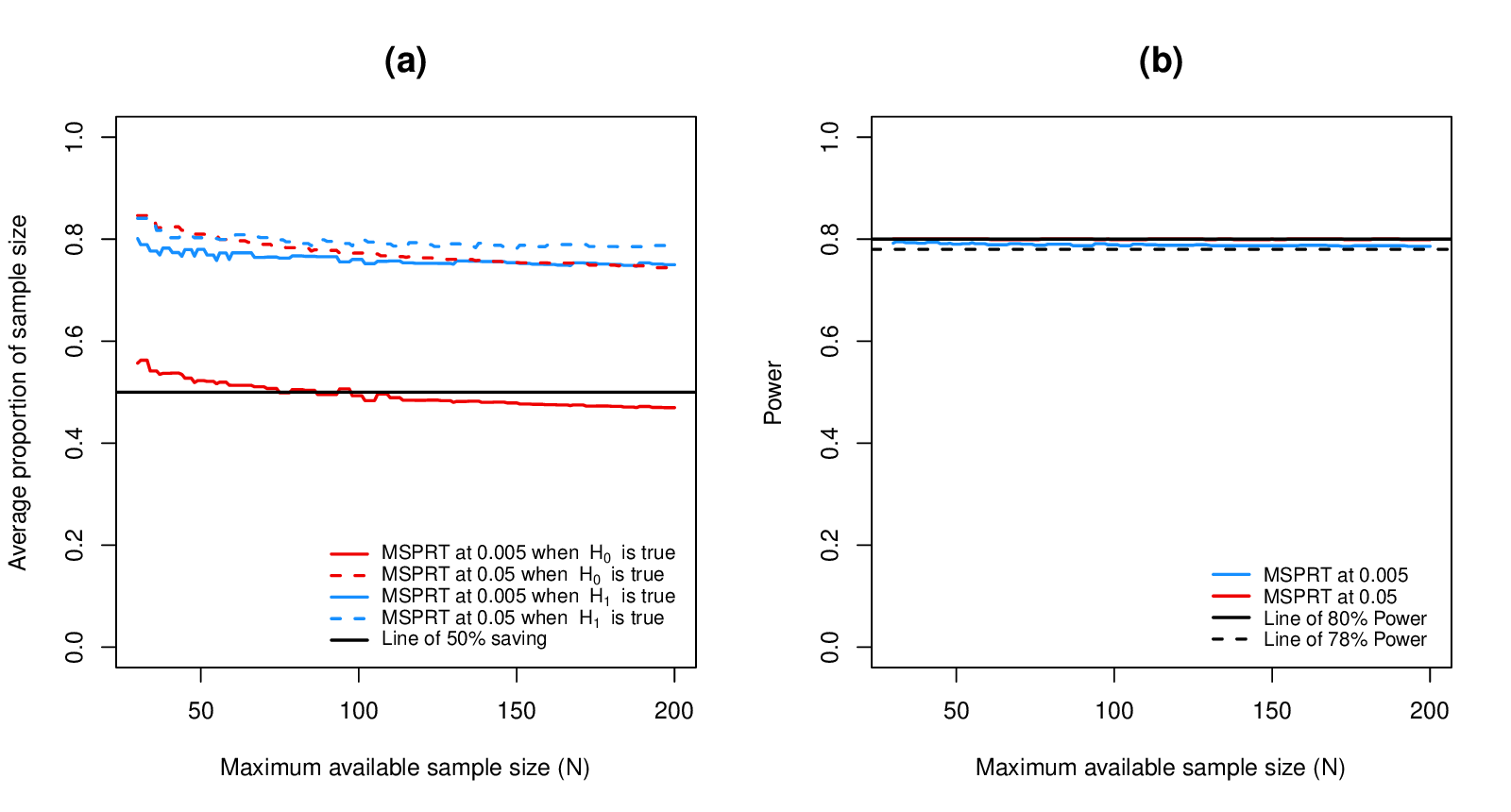}
	\caption*{{{\em Figure S8.} One-sample test that a binomial proportion equals 0.2.}  Hypothesis test of $ H_0 : \theta = 0.2$ vs. $ H_1 : \theta > 0.2 $. The curves in the left plot represent the average proportion of the maximum sample size ($N$) used before the MSPRT terminates in favor of the null or alternative hypothesis. The plot on the right displays the average power of the test against its targeted value of 0.8. In both plots, the operating characteristics under the alternative are evaluated at the corresponding fixed-design point alternatives. }\label{oneprop}
\end{figure}

Figures~\hyperlink{S6}{S6} through \hyperlink{S8}{S8} display the average proportion of the fixed-design sample size $N$ needed in a MSPRT to achieve nearly equivalent Type I and Type II errors.  In all plots, Type I errors are maintained.  The subplots on the right depict that average power achieved at the corresponding fixed-design point alternatives.

The plot provided in Figure~\hyperlink{S6}{S6} for a one-sided $z$ test is nearly indistinguishable from the corresponding plots obtained for one-sample $t$ tests and tests of a binomial proportion. For the one-sample {\em z} test and the proportion test, we get curves similar to those in Figure ~\hyperlink{S6}{S6}.  In the case of the proportion test, the discreteness of binomial data causes some non-monotonicity in the proportion of the maximum sample size that is required to reach a decision.  This feature of the plot corresponds to the non-monotonicity of power curves for fixed-design tests when sample sizes are increased.  For a given a choice of $N$, the \texttt{R} package \texttt{MSPRT} finds an ``ideal'' maximum sample size that accounts for this non-monotonicity.  We refer to these values as the ``effective sample sizes.'' In the proportion test, we illustrate the figure using only those values as the maximum sample sizes. This point is further discussed in Section~\hyperlink{S4.4}{S4.4}.

We next provide the results from simulation studies to examine the potential benefit that the MSPRT can provide in offsetting the increase of sample size that would be incurred if the bar for declaring a result ``statistically significant'' were moved from $p<0.05$ to $p<0.005$.  Specifically, we compare the sample sizes needed to achieve statistical significance at the 5\% level in standard fixed-design tests to the average sample size needed to achieve statistical significance at the 0.5\% level using the MSPRT.

From results cited in the article, this comparison is straightforward if the null hypothesis is true.  If not, care must be taken to make sure that the same alternative hypotheses are compared at both levels of significance under the fixed and MSPRT designs.  To make this comparison, we determine the $\theta^*$ that achieves the targeted Type II error in a fixed-design test of size $0.05$.  For that $\theta^*$, we next determine the $N^*$ needed to achieve the same Type II error in a fixed-design test of size $\alpha=0.005$.   We then define that $N^*$ to be the maximum sample size for the MSPRT.

\begin{figure}[h]
	\centering
	\includegraphics[width=.8\textwidth]{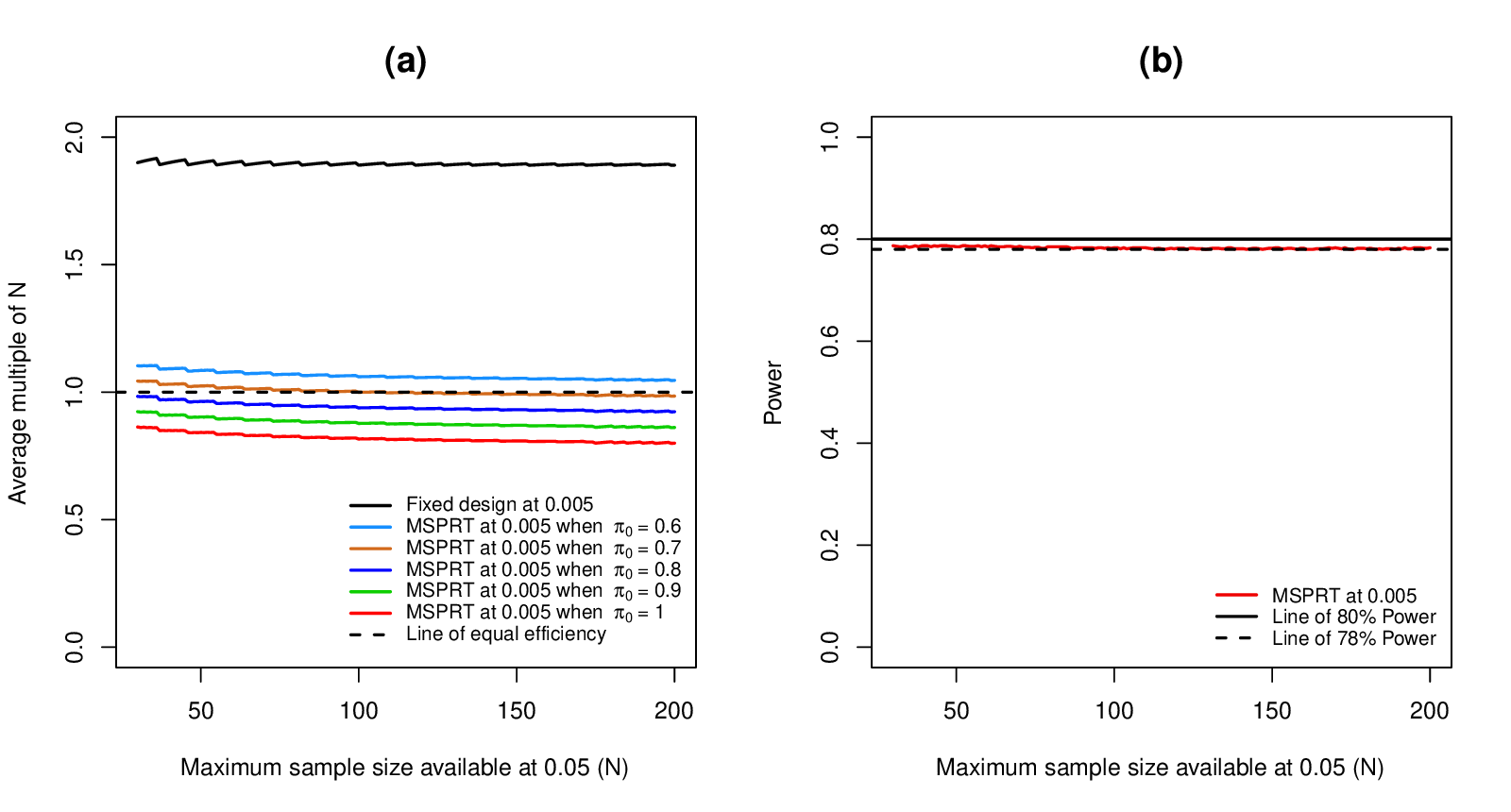}
	\caption*{{{{\em Figure S9.} One-sample}  {\em z} test that a population mean equals 0.} Curves in the left plot represent the average multiple of the sample size in a fixed-design test of size $0.05$ required in a MSPRT of size 0.005 of approximately the same power. Average sample sizes are dependent on the proportion of tested null hypotheses that are true.  The MSPRT maintains a Type I error of 0.005, and its power at $ \theta^* $ approximately equals 0.8 for the indicated proportion of $N^*$ (the sample size of the corresponding fixed-design test).  The power of the MSPRT is depicted in the plot on the right.}\label{zcomp}
\end{figure}

\begin{figure}[h]
	\centering
	\includegraphics[width=.8\textwidth]{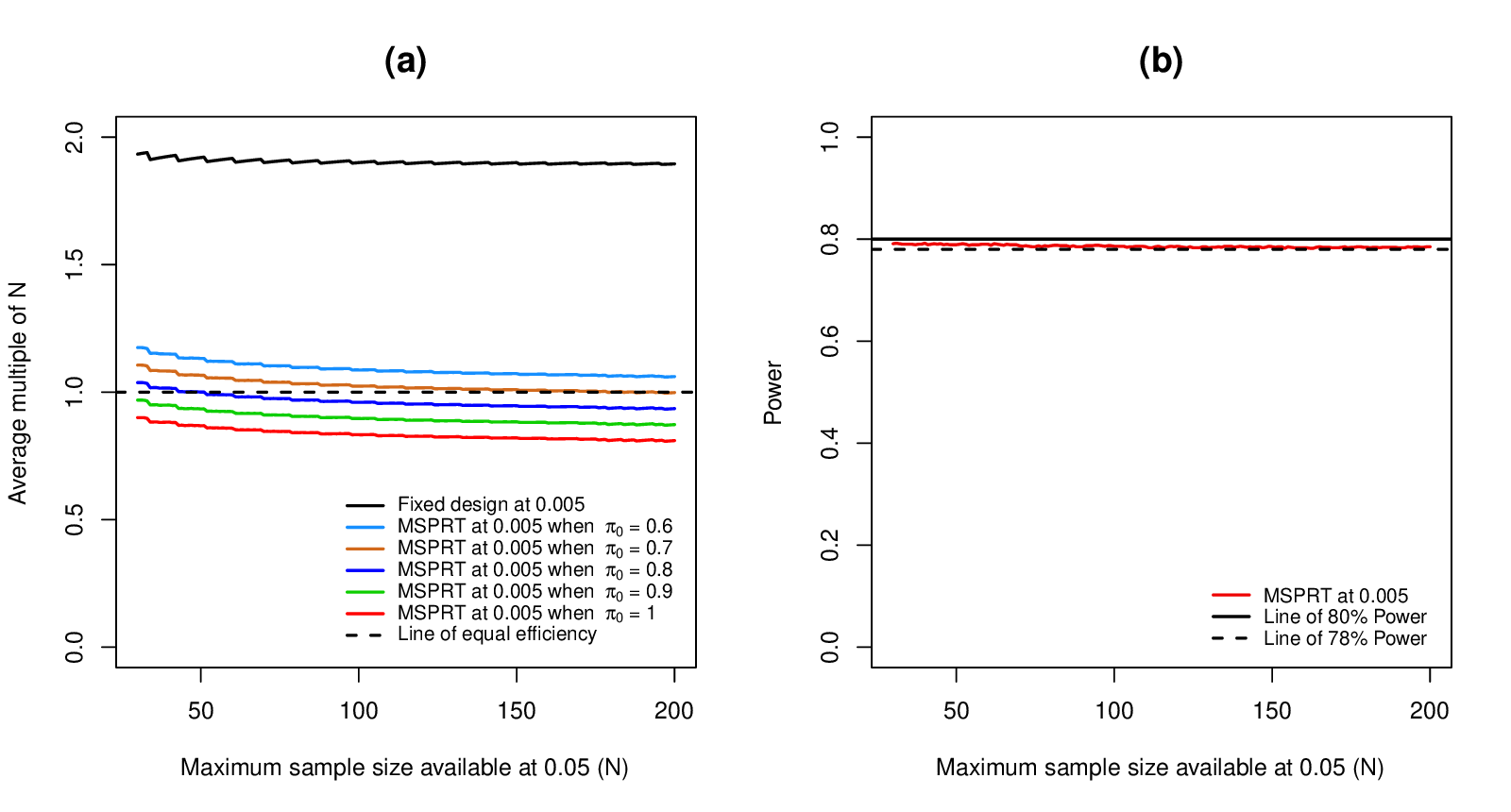}
	\caption*{{{\em Figure S10.} {One-sample} {\em t} test that a population mean is 0.} Curves in the left plot represent the average multiple of the sample size in a fixed-design test of size $0.05$ required in a MSPRT of size 0.005 of approximately the same power. Average sample sizes are dependent on the proportion of tested null hypotheses that are true.  The MSPRT maintains a Type I error of 0.005, and its power at $ \theta^* $ approximately equals 0.8 for the indicated proportion of $N^*$ (the sample size of the corresponding fixed-design test).  The power of the MSPRT is depicted in the plot on the right. }\label{tcomp}
\end{figure}

\begin{figure}[h]
	\centering
	\includegraphics[width=.8\textwidth]{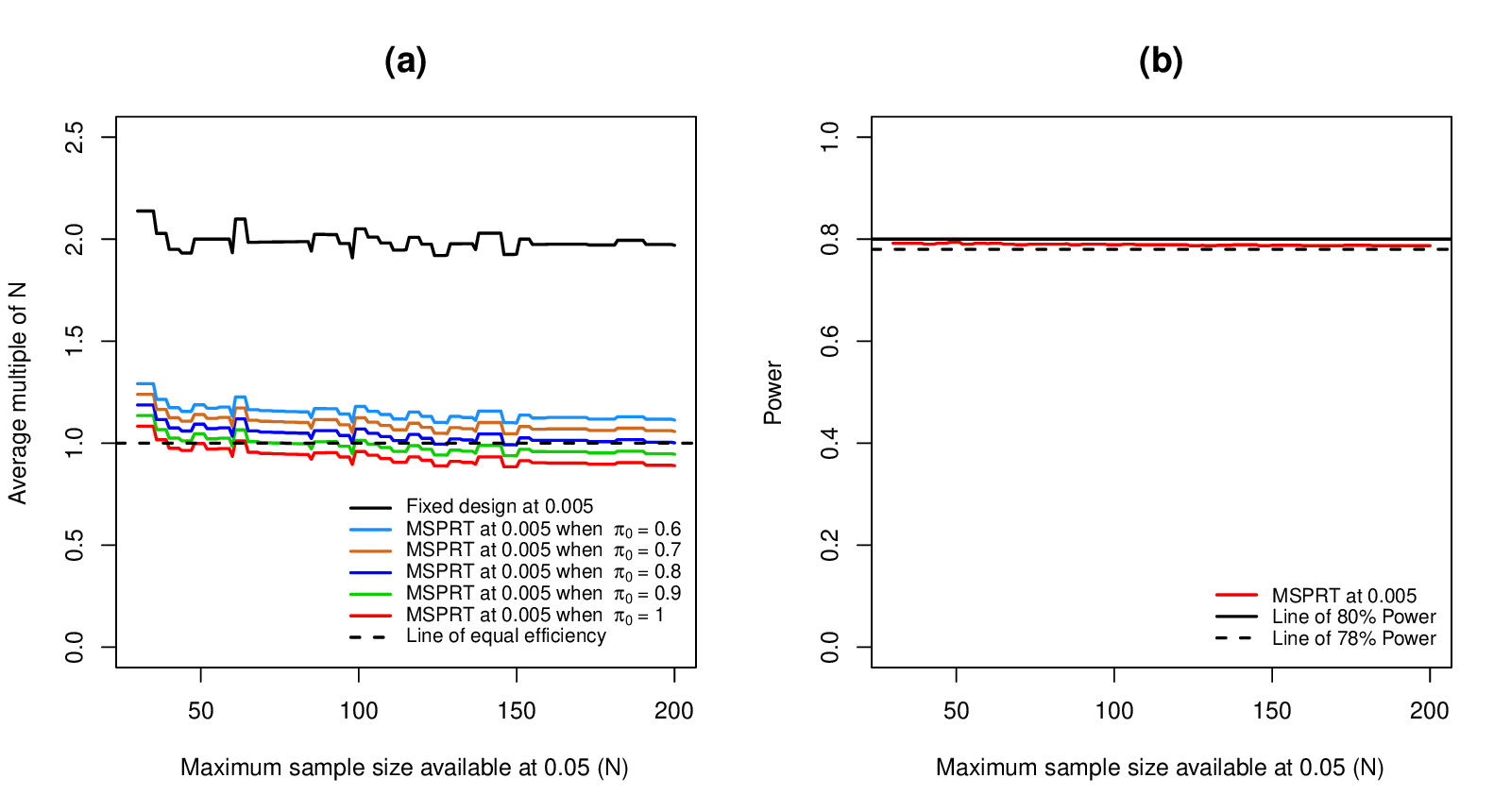}
	\caption*{{{\em Figure S11.} {One-sample} test that a binomial proportion equals 0.2.} Curves in the left plot represent the average multiple of the sample size in a fixed-design test of size $0.05$ required in a MSPRT of size 0.005 of approximately the same power. Average sample sizes are dependent on the proportion of tested null hypotheses that are true. This proportion ($ \pi_0 $) is coded by color, as indicated. The MSPRT maintains a Type I error of 0.005, and its power at $ \theta^* $ approximately equals 0.8 for the indicated proportion of $N^*$ (the sample size of the corresponding fixed-design test).  The power of the MSPRT is depicted in the plot on the right.      
}\label{propcomp}
\end{figure}


}

\medskip
\subsection{Computing the UMPBT alternative}

The UMPBT alternative is a key component of the MSPRT design. In this section we illustrate how this alternative can be obtained using the \texttt{R} package.

\medskip
\subsubsection{The $z$ test for a population mean}

Consider the test in Section~\hyperlink{S4.1.1}{S4.1.1}. To find the UMPBT alternative for a {\em z} test,  we can use the function \texttt{UMPBT.alt()}.  This command is executed as follows:

\medskip
{\small
	\begin{verbatim}
	> UMPBT.alt(test.type = "oneZ", theta0 = 3, N = 30, Type1 = 0.005, sigma = 1.5)
	[1]  3.7054
	\end{verbatim}
}

\medskip
\subsubsection{The {\em t} test for a population mean}

Similar to the {\em z} test, the function \texttt{UMPBT.alt()} also calculates the alternative for a {\em t} test. From (\ref{eq12}), it follows that the alternative is data-dependent.  Thus, we need to compute the UMPBT alternative after acquiring each data point. In order to do that, we need to specify either the sequentially observed data or the standard deviation (i.e., $s = \sqrt{\sum(x_i-\bar{x})^2/(n-1)}$ ) of the data.

Consider again the test in Section~\hyperlink{S4.1.2}{S4.1.2} with data \texttt{x}:

\medskip
{\small
	\begin{verbatim}
	> set.seed(1)
	> x = rnorm(n = 30, mean = 5, sd = 1.5)
	\end{verbatim}
}
Suppose we want to find the UMPBT alternative after observing the fifth data value.  We then need to specify either the data \texttt{x[1:5]} or the standard deviation (sd) of these data, which is roughly 1.44, in \texttt{UMPBT.alt()}. The required commands are as follows:

\medskip
{\small
	\begin{verbatim}
	> UMPBT.alt(test.type = "oneT", theta0 = 3, N = 30, Type1 = 0.005, obs = x[1:5])
	[1] 3.725457
	>
	> sd(x[1:5])   ##sd of the data x[1:5]
	[1] 1.441559
	>
	> UMPBT.alt(test.type = "oneT", theta0 = 3, N = 30, Type1 = 0.005, 
	            sd.obs = 1.441559)
	[1] 3.725457
	\end{verbatim}
}

\medskip
\subsubsection{Test for a binomial proportion}

In Table 1 of the main article we mentioned that the UMPBT alternative used by the MSPRT is a mixture distribution of two points. The function \texttt{UMPBT.alt()}  numerically computes this mixture. For illustration, consider the testing problem in Section \hyperlink{S4.1.3}{S4.1.3}. We calculate the alternative for this case with the following command:

\medskip
{\small
	\begin{verbatim}
	> UMPBT.alt(test.type = "oneProp", theta0 = 0.2, N = 30, Type1 = 0.005)
	$theta
	[1] 0.3666727 0.4000178
	
	$mix.prob
	[1] 0.2959777 0.7040223
	\end{verbatim}
}

From the output, we see that the UMPBT alternative is a mixture distribution of the two points 0.3667 and 0.4 with probabilities 0.296 and 0.704, respectively. This output corresponds to the solutions of (\ref{eq22}) and the value of $ \psi $ defined in (\ref{eq21}). The alternative illustrated above is a slight modification of what is originally defined as the UMPBT point alternative in \cite{johnson13_2}. Note that the original alternative is always the second component (\texttt{theta[2]} in the previous output) of the UMPBT alternative used by the MSPRT. This output corresponds to the solution of (\ref{eq20}).\\

{

\medskip
\subsubsection{Two-sample {\em z} test for a difference in two population means}

We again consider the testing problem in Section~~\hyperlink{S4.1.4}{S4.1.4}. To find the UMPBT alternative for a two-sample {\em z} test, we can similarly use the function \texttt{UMPBT.alt()}.  This command is executed as follows:

\medskip
{\small
	\begin{verbatim}
	> UMPBT.alt(test.type = "twoZ", N1 = 30, N2 = 30, Type1 = 0.005, 
	            sigma1 = 1.5, sigma2 = 1.5)
	[1] 0.9976144
	\end{verbatim}
}

\medskip
\subsubsection{Two-sample {\em t} test for a difference in two population means}

Similar to the two-sample {\em z} test, the function \texttt{UMPBT.alt()} calculates the alternative for a two-sample {\em t} test. From \cite{johnson13_1}, it follows that the alternative is data-dependent.  Thus, we need to compute the UMPBT alternative after acquiring each data point from both groups. In order to calculate this, we need to specify either the sequentially observed data from two groups or the estimated pooled standard deviation.

We again consider the testing problem in Section~~\hyperlink{S4.1.5}{S4.1.5} with data \texttt{x1} and \texttt{x2}:

\medskip
{\small
	\begin{verbatim}
	> set.seed(1)
	> x1 = rnorm(n = 30, mean = 0.998, sd = 1.5)
	> x2 = rnorm(n = 30, mean = 0, sd = 1.5)
	\end{verbatim}
}
Suppose we want to find the UMPBT alternative after observing the fifth observation from each group.  We then need to specify either the data (\texttt{x1[1:5]} and \texttt{x2[1:5]}) itself or the estimated pooled standard deviation of these data, which is roughly 1.005, in \texttt{umpbt.twoT()}. The commands are as follows:

\medskip
{\small
	\begin{verbatim}
	> UMPBT.alt(test.type = "twoT", N1 = 30, N2 = 30, Type1 = 0.005, 
	            obs1 = x1[1:5], obs2 = x2[1:5])
	[1] 1.004799
	>
	> sqrt(((5-1)*var(x1[1:5]) + (5-1)*var(x2[1:5]))/(5+5-2))  ## estimated pooled sd
	[1] 1.461191
	>
	> UMPBT.alt(test.type = "twoT", N1 = 30, N2 = 30, Type1 = 0.005, pooled.sd = 1.461191)
	[1] 1.004799
	\end{verbatim}
}

}

\medskip
\subsection{Obtaining the ``effective sample size" in a proportion test}\hypertarget{S4.4}{}

Because of the discreteness issue in a proportion test, power does not increase monotonically with $N$ when Type I error is exactly maintained.  We recommend choosing $N$ to make the expected sample size as small as possible. To accomplish this, a function named \texttt{effective.N()} is defined in the MSPRT package.

To illustrate this function, suppose we want to test $ H_0: p=0.2 $ against $H_1: p>0.2$ at $ \alpha = 0.005 $ with at most 30 samples. Given this choice of $ N=30 $, we  use \texttt{effectiveN.oneProp()} to determine the maximum sample size that should be used in designing the MSPRT. The command to do this is as follows:

\medskip
{\small
	\begin{verbatim}
	> effectiveN.oneProp(N = 30, theta0 = 0.2)
	[1] 28
	\end{verbatim}
}

From the output, we see that the recommended design is based on $ N=28 $ rather than $ N=30 $. If \texttt{plot.it = T} (the default), the call to \texttt{effective.N()} also returns a plot similar to that depicted in Figure~\hyperlink{S11}{S11}. This  plot shows the way an efficient $  N $ is chosen, based on decreasing point UMPBT alternatives. The green circled points correspond to the possible choices of $ N $. The largest  is chosen as the ``effective'' $  N $.

\begin{figure}[h]\hypertarget{S11}{}
	\centering
	\includegraphics[width=0.8\textwidth]{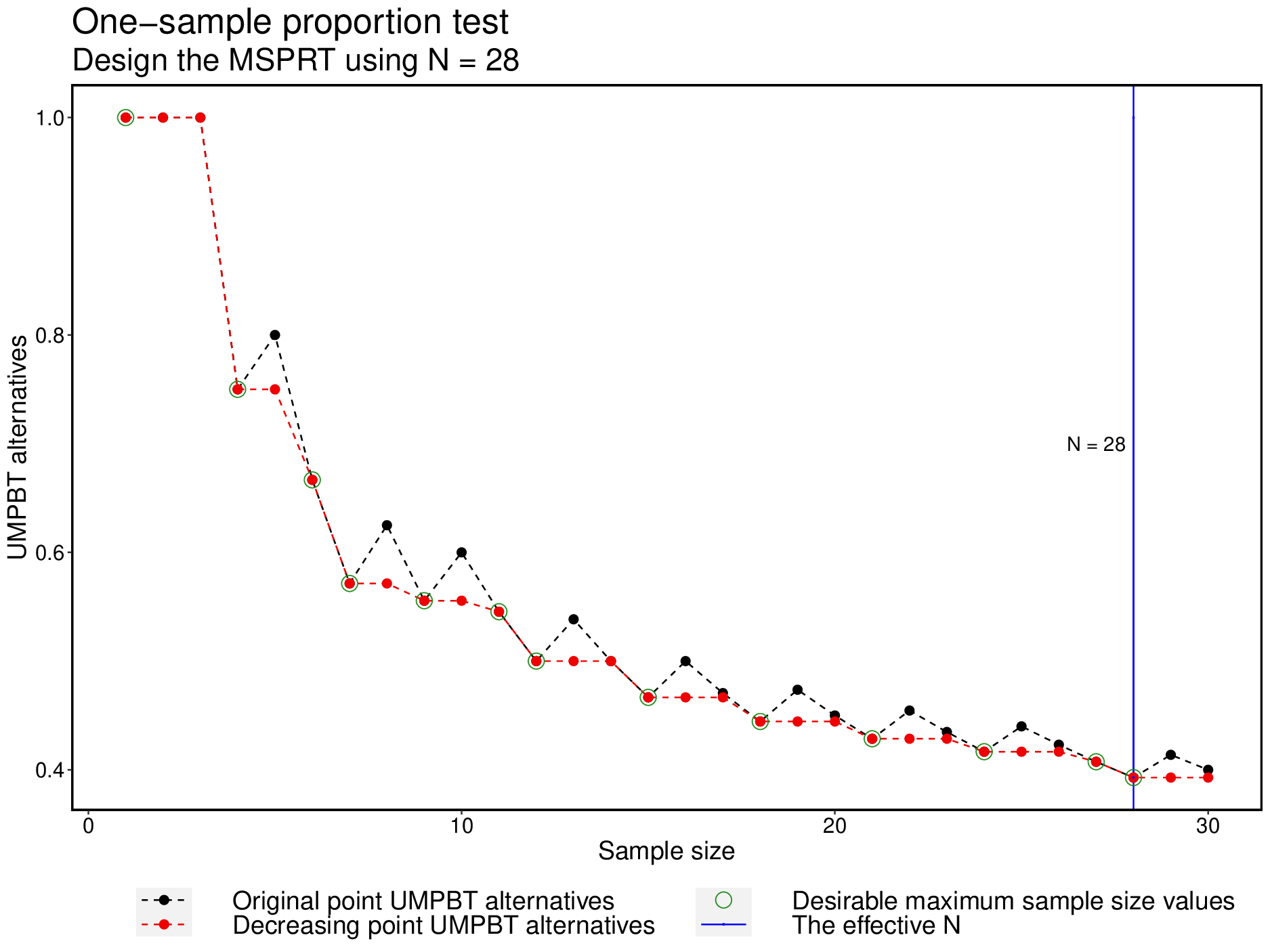}
	\caption*{{\em Figure S11.} The ``effective" $ N $ for testing $ H_0: p=0.2 $ at $ \alpha = 0.005 $.}
	\label{effectiveN}
\end{figure}

\medskip
\subsection{Finding $ N^* $}

In the main article we compared tests conducted at two levels of significance, 0.05 and 0.005. The comparison was based on the number of samples needed to achieve the higher significance level while still maintaining a prespecified power for the fixed point alternative. In those comparisons we set the point alternative to be the fixed-design alternative for the 5\% test.

To determine the fixed-design alternative in a {\em z} test for testing $ H_0: \mu = 0 $ with known $ \sigma = 1 $, $ \alpha=0.05 $ and $\beta = 0.2$, the following command can be used:

\medskip
{\small
	\begin{verbatim}
	> fixed_design.alt(test.type = "oneZ", theta0 = 0, sigma = 1, N = 30, 
	                   Type1 = 0.05, Type2 = 0.2)
	[1] 0.4539661
	\end{verbatim}
}

Now consider finding $ N^* $.  Suppose we know $ N $ for the 5\% test, and  we want at least $ 80\% $ power (the default) at the fixed-design alternative with $ \alpha=0.05 $ (that is, at 0.454 for the {\em z} test described as above). Given these constraints, the function \texttt{find.samplesize()} defined in the MSPRT package finds the required $ N^* $. In this case, the increased sample size in the {\em z} test for $ N=30 $ for the MSPRT of size 0.5\% can be found using the following command:

\medskip
{\small
	\begin{verbatim}
	> Nstar(test.type = "oneZ", N = 30)
	[1] 57
	\end{verbatim}
}

The output reveals that we need 57 samples, about twice the value of $N$, to achieve the higher significance level of 0.005 while maintaining approximately the same $ 80\% $ power at the alternative 0.454. If \texttt{plot.it = T} (the default), the call to \texttt{find.samplesize()} also returns a plot similar to that depicted in Figure~\hyperlink{S12}{S12}. This plot shows that we at least need 57 samples (red point) to meet our requirements.

\begin{figure}[h]\hypertarget{S12}{}
	\centering
	\includegraphics[width=0.8\textwidth]{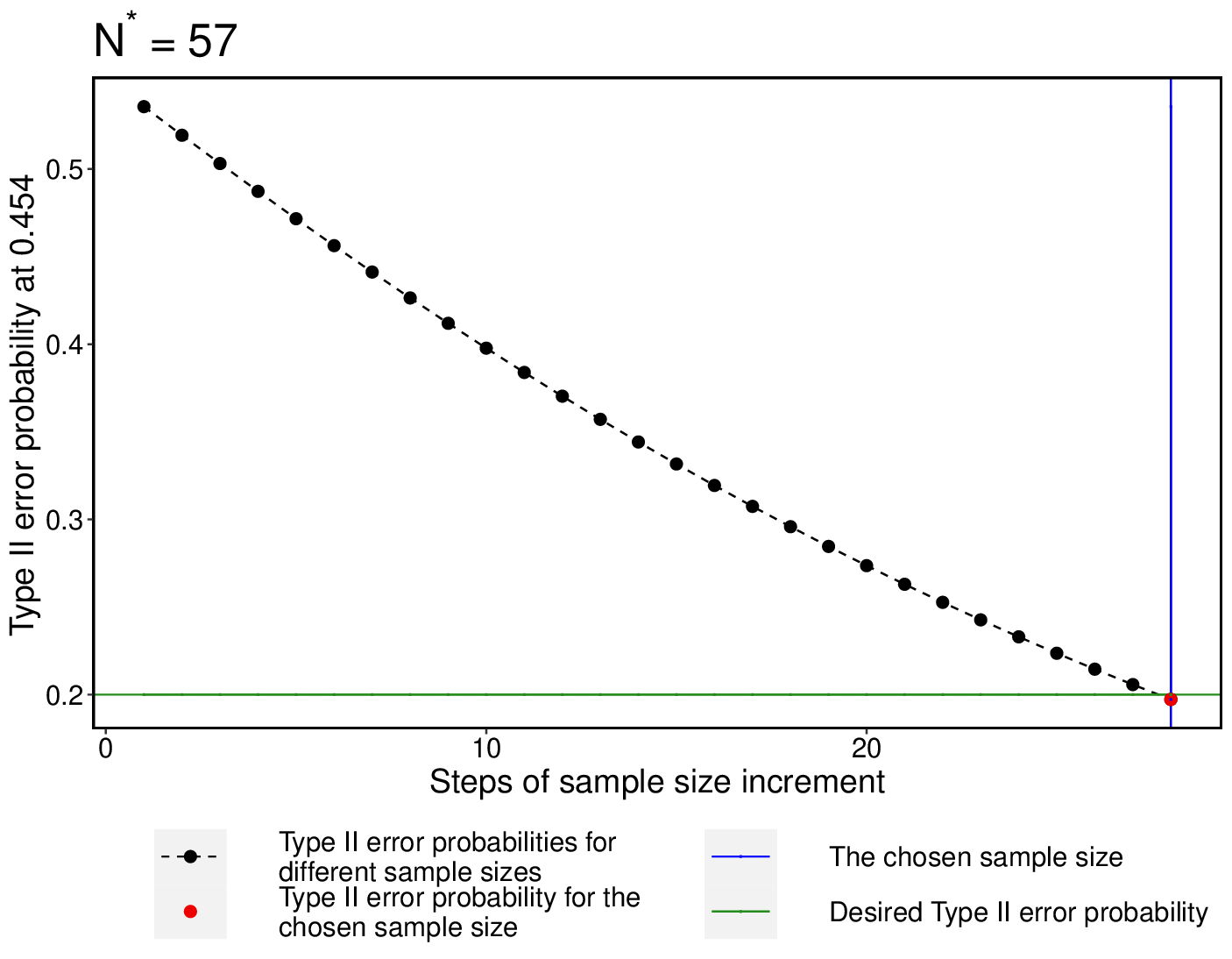}
	\caption*{{\em Figure S12.} Finding $ N^*.$}
	\label{findnstar}
\end{figure}

\bibliographystyle{apalike}
\bibliography{ref}

	
	
	